\documentclass[12pt]{aastex}
\begin{document}

\newcommand{\lap}{$L_{38}^{-1/3}$}
\newcommand{\ergs}{\rm \su  erg \su s^{-1}}
\newcommand{\etal}{ {\it et al.}}
\newcommand{\porb}{ P_{orb} }
\newcommand{\Po}{$ P_{orb} \su$}
\newcommand{\pdot}{$ \dot{P}_{orb} \,$}
\newcommand{\pot}{$ \dot{P}_{orb} / P_{orb} \su $}
\newcommand{\mm}{$ \dot{m}$ }
\newcommand{\mdot}{$ |\dot{m}|_{rad}$ }
\newcommand{\myr}{ \su M_{\odot} \su \rm yr^{-1}}
\newcommand{\msol}{\, M_{\odot}}
\newcommand{\ppp}{ \dot{P}_{-20} }
\newcommand{\cms}{ \rm \, cm^{-2} \, s^{-1} }
\newcommand{\pdott}{ \left( \frac{ \dot{P}/\dot{P}_o}{P_{1.6}^{3}}
\right)}

\def\p{\phantom{1}}
\def\pmu{\mox{$^{-1}$}}
\def\ApJ{{\it Ap.\,J.\/}}
\def\ApJL{{\it Ap.\,J.\ (Letters)\/}}
\def\ApJS{{\it Astrophys.\,J.\ Supp.\/}}
\def\AJ{{\it Astron.\,J.\/}}
\def\AA{{\it Astr.\,Astrophys.\/}}
\def\AAL{{\it Astr.\,Astrophys.\ Letters\/}}
\def\AAS{{\it Astr.\,Astrophys.\ Suppl.\,Ser.\/}}
\def\MN{{\it Mon.\,Not.\,R.\,Astr.\,Soc.\/}}
\def\Na{{\it Nature \/}}
\def\SAIt{{\it Mem.\,Soc.\,Astron.\,It.\/}}
\def\kms{km^s$^{-1}$}
\def\sbu{mag^arcsec${{-2}$}}
\def\e{\mbox{e}}
\def\dex{\mbox{dex}}
\def\L{\mbox{${\cal L}$}}
\def\gte{\lower 0.5ex\hbox{${}\buildrel>\over\sim{}$}}
\def\lte{\lower 0.5ex\hbox{${}\buildrel<\over\sim{}$}}
\def\loe{\lower 0.6ex\hbox{${}\stackrel{<}{\sim}{}$}}
\def\goe{\lower 0.6ex\hbox{${}\stackrel{>}{\sim}{}$}}

\def\lessim{\lower 0.5ex\hbox{${}\buildrel<\over\sim{}$}}
\def\gtrsim{\lower 0.5ex\hbox{${}\buildrel>\over\sim{}$}}

\def\soe{\lower 0.6ex\hbox{${}\stackrel{=}{\sim}{}$}}

\def\sgr{SGR~1900+14 }
\def\sgrp{SGR~1900+14}

\title{
The Giant Flare of 1998 August 27 from SGR 1900$+$14: \break
I. An Interpretive Study of BeppoSAX and {\it Ulysses} Observations
}

\author{M. Feroci\altaffilmark{1},
K. Hurley\altaffilmark{2},
R.C. Duncan\altaffilmark{3},
C. Thompson\altaffilmark{4}
}
\vskip 1in

\altaffiltext{1}{Istituto di Astrofisica Spaziale, C.N.R.,
  Area di Ricerca Tor Vergata, Via Fosso del Cavaliere 100,
  00133 Roma, Italy}

\altaffiltext{2}{University of California, Space Sciences Laboratory,
Berkeley, CA 94720-7450}

\altaffiltext{3}{University of Texas, Department of Astronomy, Austin,
TX 78712, USA}

\altaffiltext{4}{University of North Carolina, Department of Physics
and Astronomy, Philips Hall, Chapel Hill, NC 27599-3255}

\begin{abstract}

The giant flare of 1998 August 27 from SGR 1900+14 was extraordinary
in many ways: it was the most intense flux of gamma rays ever
detected
from a source outside our solar system; it was longer than any
previously
detected burst from a soft gamma repeater (SGR) in our Galaxy by more
than an order of magnitude; and it showed a remarkable four-peaked,
periodic pattern in hard X-rays with the same rotation
period that was found modulating soft X-rays from the star in
quiescence. 
The event was detected by several gamma-ray experiments in space,
including
the \it Ulysses \rm  gamma-ray burst detector and the BeppoSAX Gamma Ray
Burst Monitor. These instruments operate in different energy ranges,
and comparisons of their measurements reveal complex patterns of
spectral
evolution as the intensity varies.  In this paper, we present a joint
analysis
of the BeppoSAX and \it Ulysses \rm  data and discuss some implications
of these results for the SGRs.  We also present newly-analyzed 
Venera/SIGNE and ISEE-3 data
on the 1979 March 5 giant flare from an SGR in the Large Magellanic
Cloud (SGR 0526-66), and compare them with the August 27 event.  
Our results are 
consistent with the hypothesis that giant flares are due to 
catastrophic magnetic instabilities in highly magnetized neutron stars, 
or ``magnetars".  In particular, observations indicate that the initial hard
spike involved a relativistic outflow of pairs and hard
gamma rays, plausibly triggered by a large propagating fracture in 
the crust of a neutron star with a field exceeding 10$^{14}$ Gauss.   
Later stages in the light curve are accurately fit by a model 
for emission from the envelope of a magnetically-confined 
pair-photon fireball, anchored to the surface of the rotating star, 
which contracts as it emits X-rays and then evaporates completely in 
a finite time.  The complex four-peaked shape of the 
light curve likely provides the most direct evidence known for 
a multipolar geometry in the magnetic field of a neutron star.
\end{abstract}

Subject Headings:
{gamma rays: bursts -- stars: neutron -- X-rays: stars}

\section{Introduction}

During recent years, soft X-ray observations
of the quiescent counterparts to SGR 1806-20 and SGR 1900+14
have revealed periodicities in the 5--8~s range and
spindown rates of $\sim 10^{-11}$--$10^{-10}$ s s$^{-1}$
(Kouveliotou et al. 1998a, Hurley et al. 1999c).
These  rotation periods are similar to that of SGR 0526-66,
which displayed an 8~s periodicity during
the giant flare of 1979 March 5 (Mazets et al.~1979b, Barat et al.~1979).
Precise localizations of SGRs indicate that 
they may be associated with supernova remnants (SNRs) of ages
$\loe 10^4$ years.  
SGR 1806--20 was localized by detecting its bursts in X--rays
(Murakami et al.~1994) to a  position  
consistent with a radio synchrotron nebula that may be a plerionic SNR
(Kulkarni \& Frail 1993); however a recent more accurate localization of
the bursting source provides evidence that the SGR is
displaced from the radio nebula core (Hurley et al. 1999d). 
SGR 1627-41, discovered in 1998 (\cite{kouveliotou98c,feroci98}), 
is also positionally coincident with a
young SNR (Hurley et al.~1999e, Woods et al.~1999a, Smith, Bradt \&
Levine 1999). However, deep X--ray observations of SGR 1627-41
(Hurley et al. 2000) have so far failed to verify any X--ray
periodicity
in the quiescent source.

  The association of SGRs with young SNRs suggests that they are
young neutron stars. The long rotation periods and rapid
spindown rates, in the absence of any evidence for binary
companions, can be accounted for if the spindown is driven by an
ultra-strong magnetic field, e.g., $B_{dipole}\goe10^{14}$~Gauss, as
invoked in the magnetar model (Duncan \& Thompson 1992, ``DT92";
Paczy\'nski 1992; Thompson \& Duncan 1995, ``TD95").   Magnetars,
or ``magnetically-powered neutron stars,"  could form via
an $\alpha$-$\Omega$ dynamo action in hot,
nascent neutron stars if they are born spinning rapidly enough
(DT92; Thompson \& Duncan 1993, ``TD93").  Magnetism may be strong
enough within these stars to evolve diffusively over SGR
lifetimes of $\sim 10^4$ years, driving internal heat dissipation that
would keep the neutron stars
hot and X-ray bright (Thompson \& Duncan 1996; Heyl \& Kulkarni 1998). 
Above a flux density of $\sim 10^{14}$ G,
the evolving field inevitably 
induces stresses in the solid crust that cause it to yield or fracture (TD95).
Such magnetically-driven starquakes may account
for the ``ordinary" SGR bursts, which have many statistical properties
in common with earthquakes (Cheng et al.~1996; Palmer 1999;
Gogus et al.~1999;  Gogus et al.~2000).   
Less common and more catastrophic magnetic instabilities, perhaps 
involving large propagating fractures,
are thought to produce giant flares (TD95).  The non-thermal persistent
emission of the SGRs has also been proposed to be a consequence of 
magnetic activity:  either through persistent fracturing of the crust
driven by the Hall electric field (Thompson \& Duncan 1996) or through
persistent magnetospheric currents that are excited by twisting motions
of the crust during outbursts (Thompson et al.~1999).

We note that the observed spindown histories of SGRs are 
{\it not} well-fit by the idealization of vacuum magnetic dipole 
radiation [MDR] (Kouveliotou et al.~1998a, 1999; Marsden et al.~1999; 
Woods et al.~1999b, 2000). Indeed,
spindown torques in active magnetars can be enhanced by persistent
outflows of relativistic particles and Alfv\'en waves, channeled
by a strong magnetic field (Thompson \& Blaes 1998; 
Harding, Contopoulos \& Kazanas 1999; Thompson et al.~1999) and also 
strongly modulated by material ejected during bursting activity
(Woods et al.~2000).    
It remains possible that the smooth spindown in some {\it inactive} 
candidate magnetars (such as 1E 1841-045) is well fit by MDR
(Vasisht \& Gotthelf 1997; Gotthelf, Vasisht \& Dotani 1999).
The observed spindown in SGRs is consistent with dipole fields weaker than
the QED strength ($4.4\times 10^{13}$ G), but only if a persistent
outflow of particles carries away substantially more energy
than is observed in X-rays  (Harding, Contopoulos \& Kazanas 1999; 
Marsden et al.~1999).

Accretion-powered alternatives to the magnetar model have 
also been considered 
(e.g., Van Paradijs, Taam \& Van den Heuvel 1995;  Li 1999; Chatterjee, 
Hernquist \& Narayan 2000; Alpar 2000).  Such models make 
reasonable fits to the 
continuous X-ray emissions and spindown histories of SGRs and Anomalous 
X-ray Pulsars (AXPs), but they offer no good explanation for the 
hyper-Eddington burst and flare emissions that are the defining property
of SGRs (cf.~\S 7.3 in TD95; \S 5.2 in Thompson et al.~1999).  
 
This paper will focus on SGR 1900+14.  Of the four known SGRs, this
object is the only one without an identified  SNR 
surrounding it. However, the recently-verified source position
(Vasisht et al.~1994; Hurley et al.~1999a; Murakami et al.~1999; Frail,
Kulkarni \& Bloom 1999)
lies just outside the edge of G42.8+0.6, a $\sim 10^{4}$-year-old
galactic SNR. \ A parallel can be drawn with the other giant flare source,
SGR 0526-66, which lies near the edge of, but just
{\it inside}, the young SNR N49 in the LMC (Cline et al.~1982). 
Note that several mechanisms could plausibly impart a recoil velocity
$\sim 1000$ km s$^{-1}$ to a magnetar at birth, sufficient to propel
it outside its remnant in $\sim 10^4$ years (DT92).

SGR 1900+14 was first detected in 1979, when it emitted three bursts
of soft gamma-rays of moderate intensity (Mazets et al.~1979a).
The source again became burst active for a short while
in 1992 when a handful of events were recorded (Kouveliotou 1993).  In 1998
May, SGR 1900+14 entered an unprecedented level of burst activity in both
frequency and intensity (Kouveliotou et al.~1998b, Hurley et al.~1999a).  
Purely by chance, an ASCA (2--10 keV) observation less than one
month prior to the source reactivation in 1998 led to the discovery of 5.16
pulsations from the quiescent counterpart to the SGR (Hurley et al. 1999c).
Subsequent RXTE PCA (2--20 keV)
observations following the reactivation of the SGR allowed
for a measurement of the period derivative 
$\sim 1 \times 10^{-10} \rm s \, s^{-1}$ (Kouveliotou et al.  1999; 
Woods et al. 1999b).
This spindown rate is consistent with a magnetar-strength field
(Kouvelioutou et al.~1999; Thompson et al.~1999 and references therein).

 On 1998 August 27 a giant flare from SGR1900+14, lasting more than
five minutes, was detected by Konus-Wind, \it Ulysses \rm  and BeppoSAX
(Cline et al. 1998; Hurley et al. 1999b; Feroci et al. 1999; 
Mazets et al.~1999).
Gamma rays during the first second were extraordinarily intense,
overwhelming detectors on several other spacecrafts as well, 
such as the Near Earth Astreroid Rendevous mission (which went into
a protective shut-down mode) and 
the Proportional Counter Array on the Rossi X-ray Timing Explorer.  
The Compton Gamma-Ray Observatory was Earth-occulted for this flare.

The 5.16~s neutron star rotation period was clearly detected during the
giant flare.  Indeed, the periodic signal was intense enough to produce
a marked 5.16-second modulation in the height of the Earth's ionosphere,
which affected long-wavelength radio transmissions (Inan et al.~1999),  
a remarkable effect for a star $\sim 20,000$ light years away.   
About 40 seconds after the onset of the flare,a 1.03 s
repetitive pattern (4th harmonic of 5.16 s fundamental) was observed
to set in gradually (Feroci et al.~1999; Mazets et al.~1999), 
unlike any emission previously detected from any source.
A radio afterglow was soon found with the Very Large Array
(Frail, Kulkarni \& Bloom 1999).  This 
source, the only radio point source unambiguously associated with an SGR,
was apparent in the error box of SGR1900+14 on 1998 September 3,
but it faded away in less than one week, giving evidence for an
abrupt outflow of relativistic particles during the flare.

In this paper we carry out a comparative analysis of BeppoSAX and {\it
Ulysses} observations of the August 27th event.  Our main goal is
a thorough description of the data that may be cogent to
understanding SGR giant flares, including some novel comparisons with
the 1979 March 5th event.  
The observed properties are then qualitatively interpreted in the context 
of the magnetar hypothesis, deriving insights and constraints on the
magnetic field structure and mechanisms for large flares in SGRs.
A companion paper (Thompson et al.~2000, hereafter Paper II) derives 
a more quantitative and rigorous treatment of the physics and magnetic
field structure in this and similar objects.

\section{Observations}

\subsection{Instrumentation}

\subsubsection{BeppoSAX Gamma Ray Burst Monitor}

The GRBM consists of the anticoincidence detectors of the
Phoswich Detection System (PDS, Frontera et al. 1997),
comprised of four optically independent CsI(Na) shields forming a
square box, surrounding the main PDS detectors.
Each shield is 1 cm thick and has
a geometric area of about 1136 cm$^{2}$.  The maximum effective area for
a burst with a typical power law spectrum arriving at normal incidence
is about 420 cm$^{2}$ for unit 1 or 3 (the optimum units), when
shadowing by spacecraft and experiment structures and the detector 
response are taken into account.

The GRBM electronics records data from each shield in both real-time
(low time resolution) and triggered (high time resolution) modes.  
The real-time data consist of 1 s resolution count rates in the 
40-700 and $>$100 keV energy ranges.  
The triggered data consist of: \
(a) 7.8125 ms data for 8 s prior to the trigger time,
(b) 0.48828 ms data for 10 s starting at the trigger time, and
(c) 7.8125 ms data for 88 s starting at 10 s after the trigger time.
These count rates are all recorded in the 40-700 keV energy range.

Independent of the trigger,
256-channel energy spectra are taken every 128 s,
for each of the four shields. These are integrated over fixed time
intervals,
and are therefore mainly useful for calibration purposes. In fact, given
the typical time scales of cosmic GRBs, the housekeeping
spectral data can only be used for obtaining the average energy spectra
of bright GRBs, when they can be detected over
128 s of background.

The 1 s real-time data in the two energy channels,
40-700 keV and $>$700 keV,
allow for some spectral reconstruction. The two count rates
overlap in the nominal energy range from 100 to 700 keV. 
From these, the 40-100 keV and 100-700 keV rates may be derived,
with assumptions about the number of counts above 700~keV\footnote{
We have no definite way of knowing what the counts are
above 700 keV. However, in this specific case, 
we know from the time-averaged spectra what
the spectrum is below 700 keV and can extrapolate above this energy.
In addition, the GRBM effective area above 700 keV decreases continuously,
starting at $\sim$25\% of its maximum value at 700 keV (the
maximum is reached at $\sim$200 keV) (\cite{amati99}).
This fact, combined with a photon spectrum decreasing with
energy (in our case the spectrum is very hard only in the first seconds,
when we don't make any use of the realtime ratemeters),
results in an estimated number of counts that is negligible with
respect to the count rates in 40-100 and 100-700 keV.
For this reason we simply preferred not to make any correction, which
would be almost entirely arbitrary, and leave the results that we present
subject to this assumption.} (\cite{amati97}).
Therefore, these ratemeters can be used to extract
a 2-channel spectrum, with a time resolution of 1 s. Note that
the events that are recorded in these two ratemeters are exactly
the same in their overlapping energy range, and are therefore
completely covariant, significantly lowering the statistical
error in their difference.

The energy resolution of a shield ranges from 15\% to 30\%, depending on
the energy and position of interaction of the photon. Additional 
details on the GRBM instrument and its in-flight performance can 
be found in Feroci et al. (1997).

\subsubsection{\it Ulysses \rm  Gamma Burst Detector}

The \it Ulysses \rm GRB detector (Hurley et al. 1992) 
consists of two 3 mm thick
hemispherical CsI(Na) scintillators
with a projected area of about 20 cm$^2$
in any direction.  The detector is mounted
on a magnetometer boom far from the body of
the spacecraft, and therefore has a practically
unobstructed view of the full sky. 
Because the \it Ulysses \rm mission is in interplanetary
space, the instrument benefits from an exceptionally stable background. 
The energy range is $\sim$ 25-150 keV. 
The lower energy threshold is set by a discriminator,
and is in practice an approximate
one; photons with energies $>10$ keV can penetrate
the housing and be counted either
because of the rather poor energy resolution at
low energies, or, in the case of
very intense events, due to pulse pile-up. 
For the 1998 August 27 event, an intense
flux of low energy photons was present, and both of these effects
operated to
some extent.  The upper energy threshold is set
by a discriminator, and also by the decreasing detector efficiency. 
The instrument took time history data of the August 27 event in both 
triggered and real-time modes.  The triggered data had time resolution 
0.03125 s, but it lasted for only 64 s; the real-time data had resolution  
0.5 s, and was transmitted thoughtout the event.  The first
$\sim$7.3 s of the triggered data recorded the burst prehistory.
The instrument also generally records 16 channel energy spectra
for a total duration of $\sim$500 s after any trigger, with progressively
longer
time resolutions, starting at 1 s and ending at 16 s.  The energy
resolution is
$\sim$ 27 \% at 60 keV. 

\subsection{The August 27 event}

The giant outburst from SGR 1900+14 triggered the BeppoSAX
Gamma Ray Burst Monitor and the \it Ulysses \rm Gamma Ray Burst
detector on August 27 1998, 10:22:15.7 UT (Feroci et al. 1999,
Hurley et al 1999b).
At the time of this event, SGR 1900+14 was located at 
an elevation angle of 48$^{\circ}$ with
respect to the GRBM equatorial plane, and at an azimuthal angle
of 29$^{\circ}$ with respect to the GRBM unit 1, whose data 
will be used in the analysis presented in the following sections.
The effective area of the GRBM unit 1 at this incidence angle
varies from $\sim$56 cm$^{2}$ at 60 keV to $\sim$365 cm$^{2}$ at
280 keV. The GRBM unit 1 recorded $\sim$10$^{6}$ counts in the 
40--700~keV range and $\sim$3$\times$10$^{5}$ above 100~keV.
The peak count rate in 40--700~keV was estimated to be 
$\sim$1.5$\times$10$^{5}$ counts s$^{-1}$ but it was probably 
affected by saturation problems (Feroci et al. 1999).
The \it Ulysses \rm GRB detector recorded $\sim$1.8$\times$10$^{6}$
counts in the energy range 25-150~keV, with a peak count rate  
of $\sim$2$\times$10$^{5}$ counts s$^{-1}$, also possibly
affected by pulse pile-up and dead-time effects.

A reliable estimate of the energetics of the event is made difficult
by pile-up and dead-time problems.
Mazets et al.~(1999) made a signficant effort to derive tight lower
bounds using the Konus experiment.
For the energy range $>$15~keV, they found a lower bound on the peak flux
of $3.1\times10^{-2}$ erg cm$^{-2}$ s$^{-1}$,
and a fluence $> 5.5\times 10^{-3}$ erg cm$^{-2}$ in the hard spike
(first 0.45 s).  At a distance of $10 \, D_{10}$ kpc, and assuming
isotropic emission, this corresponds to a peak luminosity greater than
$3.7\times10^{44} \, D_{10}^2$ erg s$^{-1}$ and a total hard spike
energy $ >7\times 10^{43} \, D_{10}^2$ erg.
The fluence subsequent to the hard spike was $4.5 \times 10^{-3}$ erg
cm$^{-2}$ in $> 15 $ keV photons (Mazets et al.~1999).
A substantial fluence in lower-energy photons (Inan et al.~1999) and
in neutrinos (\S 5 in TD95; Paper II) is also likely.  Thus a lower bound on
the total event energy is $ 2\times10^{44} \, D_{10}^2$ erg.

\subsection{Available Data}

The available real-time data from \it Ulysses \rm
and BeppoSAX Gamma Ray Burst detectors consist of
count rates over 0.5~s in the 25--150~keV band (\it Ulysses),
\rm and count rates over 1~s in the 40--700, 40--100 and 100--700~keV
bands (BeppoSAX).  In both instruments these rates
are accumulated onboard over fixed time intervals.
Taking into account the location of SGR~1900+14 and the two spacecraft
positions, we find that
the onboard integration times of the two experiments are such
that the first time bin containing photons from the giant flare
starts for \it Ulysses \rm  at 37335.168 SOD (seconds of day
August 27th, 1998) and for GRBM at 37335.05653 SOD.
Therefore, purely by chance, the two low-resolution light curves
are synchronized to within approximately 100~ms 
($\sim$ 0.02 cycles of the neutron star rotation), and they can be
used for time-resolved spectral analysis.
The light curves of the event in these energy ranges are shown in
Fig.~\ref{3decays}.  The effective area
of the GRBM for this event was only $\rm \sim 56 \, cm^2$ at 60 keV
(\cite{feroci99}); because of this and the falling energy spectrum, the
numbers of counts detected by GRBM and \it Ulysses \rm are comparable.

High time-resolution data were recorded only during the first portion
of the event. In particular, the BeppoSAX/GRBM
provided 7.8125~ms data for the first $\sim$98~s of the event
in the 40-700~keV range,
while the \it Ulysses \rm  GRB detector recorded the first
$\sim$57~s of the event with a time resolution of 31.25~ms
in the nominal energy range 25-150~keV (with a likely contribution
from photons of energies $>$10~keV.

\section{Timing Analysis}

\subsection{Envelope of the Light Curve}\label{envel}

The envelope of the light curve decays smoothly, and provides an
important clue to the radiative mechanism.  
The simplest choice of an exponential function $\exp(-t/\tau)$ 
adequately describes the intermediate portion
of the decay (Fig.~\ref{3decays}).  The best fit time constant varies
slightly from $\tau = 78$~s in the \it Ulysses \rm energy range to
70~s in both the 40--100 and 100--700~keV GRBM energy ranges.
An even longer decay constant of $\sim$ 90~s was fit by
Mazets et al. (1999) in the Konus energy range ($>$15~keV), which
is softer than \it Ulysses \rm.  This decrease in the decay constant
with increasing photon energy suggests a slight
overall softening in the spectrum on a timescale of $\sim 100$ s. 

Note, however, that the data drops sharply below the exponential
fit at $\sim$300~s after the event onset.  This leads us to consider
a second parameterization motivated by a cooling fireball which
is trapped on the closed magnetic field lines of a neutron star 
(Paper II):
\begin{equation}\label{lxt}
L_X(t) = L_X(0)\left(1-{t\over t_{evap}}\right)^{a/(1-a)}.
\end{equation}
In this expression, the
cooling luminosity is assumed to vary as a power of the {\it remaining}
fireball energy, $L_X \propto E^a$, and $t_{evap}$ is the
time at which the fireball boundary propagates to its center and the
fireball evaporates.
The fireball index $a$ accounts for the geometry and the temperature
distribution of the trapped fireball, being ${2\over 3}$ for a spherical 
trapped fireball of uniform temperature. 
Giant outbursts such as the August 27 event correspond to fireball
interior temperatures of $T \sim 1$ MeV if essentially all the burst energy is
released during the initial spike, and a sizeable fraction of this
energy is trapped (TD95).  In this situation, the scattering
depth across the fireball is so large, $\tau_{es} \sim 10^{10}$, that
the fastest mode of radiative loss involves the inward propagation
of the cool boundary of the fireball.  Neutrino pair cooling causes
significant deviations from eq. (\ref{lxt}) above $T \sim 1$ MeV, as we
discuss in Paper II.

The function (\ref{lxt}) is fit to the \it Ulysses \rm data in
Figs. \ref{fire_ul_reb} (rebinned at 5-s intervals, the closest
available approximation to the 5.16-s period, in order to reduce
the scatter caused by the oscillations) and \ref{fire_3decays}a.
In order to fit the 5-s \it Ulysses \rm lightcurve we selected 
the time interval going from 50 to 450~s after the event onset,
that is from when the large amplitude oscillations set in up to 
when the event is finished.
A fit to the whole light curve with all the three parameters of the
cooling fireball model free to vary brings to the following set of values:
[$L_{X}(0)$=(64,308$\pm$279) counts/5s, $t_{evap}$=(525$\pm$5)~s
and $a = 0.821 \pm 0.002$], with a $\chi ^{2}$=5132 (78 degrees of 
freedom). (All errors are given at 1-$\sigma$ significance level). 
Since the physical meaning of $t_{evap}$ in the cooling trapped fireball
model is the time at which the fireball itself evaporates 
(corresponding to the time at which the light curve goes to the baseline),
we evaluate $t_{evap}$=525~s unacceptable. Most likely, the 
reason for this result is that the best statistics 
(which drives the $\chi^{2}$ minimization) for the lightcurve is
available in the first portion of the curve, whereas $t_{evap}$
is mostly characterized by the tail of the curve itself.
Therefore, we used the final portion of the curve, from t=250~s to
t=450~s, where there is not a large variation in the statistical 
quality of the data, to derive a more reliable value for $t_{evap}$.
In fact, fitting this final portion of the curve we derive:
[$L_{X}(0)$=(57,913$\pm$3644) counts/5s, $t_{evap}$=(411$\pm$3)~s
and $a = 0.742 \pm 0.005$], with $\chi^{2}$=251 (38 d.o.f.).
Then, we go back and consider again the curve starting at t $>$ 50~s,
and fit the model to the data, allowing $t_{evap}$ to vary only
within the 1-sigma range provided by the fit to the t$>$50~s curve.
With this procedure we derive the following best-fit parameters:
[$L_{X}(0)$=(58,513$\pm$165) counts/5s, $t_{evap}$=414~s (at the limit)
and $a = 0.756 \pm 0.003$], with a $\chi^{2}$=6086 (79 d.o.f.).
The excellence of the fit is demostrated by the dashed, coloured
curves in Fig.~\ref{fire_ul_reb}, showing how the trapped
fireball model follows the decay trend of the curve, and in particular
accurately matches the sudden final drop in flux
(except for the green curve, where $t_{evap}$ appears clearly
overestimated).
For indices $a$ not far from the value corresponding to a
spherical fireball of uniform temperature ($a = {2\over 3}$),
it tracks the envelope of the light curve over the entire phase of
large-amplitude pulsations (Fig. \ref{fire_ul_reb} and \ref{fire_3decays}a).
We note, however, that fixing $a = {2\over 3}$ during the fitting 
procedure provides a large increase in
the value of $\chi^{2}$ (more than a factor of 4). 
In the following we will assume the last set of trapped-fireball
parameters as best-fit parameters, however a word of caution is needed
due to the large values obtained for the $\chi^{2}$, indicating that the 
light curve has intrinsic variability, much larger than the 
statistical errors.

An indirect confirmation of the goodness of our choice is obtained when
we compare the derived fireball model to the high energy (GRBM) data
in the two energy intervals.
In Fig. \ref{fire_3decays}b-c we show the results of the fit to
the 40-100 and 100-700 keV lightcurves (for 50~s $<$ t $<$ 450~s).
The blue curves correspond to fits done with the three parameters
of the trapped-fireball model free to vary. The resulting values are:
[$L_{X}(0)$=(6,036$\pm$52) counts/s, $t_{evap}$=(501$\pm$13)~s
and $a = 0.828 \pm 0.005$] for 40-100~keV ($\chi^{2}$=867, 77 d.o.f.)
and
[$L_{X}(0)$=(1,646$\pm$53) counts/s, $t_{evap}$=(545$\pm$62)~s
and $a = 0.85 \pm 0.01$] for 100-700~keV ($\chi^{2}$=123, 77 d.o.f.).
Instead, the red curves correspond to fits done fixing $t_{evap}$
within the 1-$\sigma$ range provided by the analysis of the 25-150 keV data
and leaving $a$ and $L_{X}(0)$ free to vary. 
The best-fit parameters for these fits are:
[$L_{X}(0)$=(5,720$\pm$31) counts/s, $t_{evap}$=414 s
and $a = 0.785 \pm 0.001$] for 40-100~keV ($\chi^{2}$=938, 78 d.o.f.)
and
[$L_{X}(0)$=(1,542$\pm$30) counts/s, $t_{evap}$=414 s
and $a = 0.794 \pm 0.003$] for 100-700~keV ($\chi^{2}$=129, 78 d.o.f.).
We see that the GRBM energy ranges require a fireball index larger
than that obtained in 25-150~keV.
This behavior is consistent with a mild softening
of the spectrum during the phase of large-amplitude pulsations,
already suggested by the exponential-decay fits, which
causes the light curve to drop more rapidly at higher energies.

We note that all our favored fireball indices differ slightly, but 
significantly,
from that expected in case of spherical geometry ($a = {2\over 3}$). 
We will comment on the
physical significance of the $a$ parameter values in Paper II.
In addition, both exponential and trapped fireball fitting functions 
significantly
underestimate the measured flux during the first stages
of the August 27 outburst --  the initial hard pulse and the ensuing
smooth $\sim 40$ s decay.  This excess flux disappears just as the
large-amplitude pulsations begin to emerge.  

In Figure~\ref{smooth_decay} we present the first $\sim$100~s of the 
event, as derived from the high resolution 
GRBM lightcurve rebinned with $\sim$5.16 s resolution, i.e. one spin period.
The decay appears to be quite smooth and monotonic,
apart from a small feature around 37370 SOD, or $t = 40$ s after
burst onset, which is coincident with the first appearance of the subpulses
(see Figs.~\ref{grbm_panels} and \ref{Ulysses_grbm_diff}). 
This indicates that
the onset of the four-peaked structure in the light curve
does not correspond to an additive contribution to the emission -- with
the possible exception of a short, modest, transient enhancement around
37370 SOD -- but rather to a redistribution of the same emission over
the spin phase.  This is consistent with the idea
that a rapid release of energy occurred at the event's onset, and that the
subsequent, generally declining X-ray emissions are due largely to the
loss of this energy from the star and its vicinity.

\subsection{Power Spectra}

The real-time data can be used to characterize the energy and time
dependence of the 5.16-s period pulse shape.
In Fig.~\ref{fig_psd} we show the power spectral density (PSD) of 
the \it Ulysses \rm  and GRBM real-time data over four consecutive time 
intervals covering the entire event (from 3-50~s, 50-150~s, 150-250~s,
250-350~s after the event onset, except for 100-700~keV where we limited
the last Fourier transform to the interval 250-300~s, due to the 
low statistics). 
In order to subtract the low frequency noise introduced by the slow decay,
for each of the three energy ranges the best-fit trapped fireball model
was subtracted from the data. In the interval 3-50~s the excess of the
data with respect to the trapped fireball model motivated an 
additional detrending with a second-order polynomial form.

The many peaks apparent in the PSD are all but one 
related to the fundamental
frequency ($\sim$0.2~Hz) of the star's rotation, being either 
high order harmonics (up to n=5 for \it Ulysses \rm and n=2 for GRBM)
or the aliases of the higher order harmonics, 
due to the coarse sampling of the data.  
In particular, we expect contributions in our spectra from  
the following aliases in the \it Ulysses \rm 
data (n indicates the harmonics that
are aliased at the given frequency - we only mention up to n=10):
0.837~Hz (n=6), 0.643~Hz (n=7), 0.449~Hz (n=8), 0.255~Hz (n=9),
of which the first one is clearly detected and the other three are
possibly detected, but marginally.
For the GRBM data we expect aliases at frequencies:
0.418~Hz (n=3), 0.225~Hz (n=4), 0.449~Hz (n=8), 0.255~Hz (n=9),
of which the first one is clearly detected.

The only peak that cannot be attributed to the source is visible at 
$\sim$0.08~Hz in the PSD of the \it Ulysses \rm data.
This is an instrumental effect, related to the spin period
of the \it Ulysses \rm spacecraft (5 rpm = 0.0833~Hz),
which causes a periodic absorption
of the source flux by the carbon fiber magnetometer boom. 
The PSD obtained from BeppoSAX data do not suffer similar
problems, because the satellite is three-axis stabilized.
Other spurious modulations are possible due to 
background geomagnetic variations and Earth occultation over the 
BeppoSAX orbit, but only at frequencies $\lessim$~0.2~mHz that are
too low to be probed by this transient flare data. 

The results of the Fourier transforms contain information about how the
pulse shape varies with time and photon energy.
The fundamental pulsation is only marginally detected
during the first time interval in any of the three energy ranges.
This indicates that the light-curve is not strongly modulated on
the rotation period, as evident in the 
first $\sim 40$ s of the high time resolution data  (see 
Fig.~\ref{Ulysses_grbm_diff}). 
The absence of large-amplitude modulations suggests that an optically
thick plasma filled a larger volume of the magnetosphere;  and the
relative lack of phase coherence points to large fluctuations in the position
of the scattering photosphere, well outside the volume containing most
of the burst energy.
As the event progresses, the rotational modulation becomes more pronounced,
and the soft X-ray power in high-order harmonics 
varies significantly relative to the fundamental.   
It decreases near the end of the event, indicating that the pulse shape 
evolves toward a smoother profile.
Interestingly, the 0.837~Hz peak (alias of n=6)
in the 25-150~keV PSD increases its power with time, relative
to the fundamental frequency.  
In the last time interval (starting 250~s after the event 
onset) the PSD of the \it Ulysses \rm
data shows that most of power at frequencies above the fundamental
is contained in the $\sim$1~Hz peak.  This peak is strong 
likely due to a four-peaked structure of the light curve 
which is quasi-repetitive on a 1-s period,
as it is observed in the early phases of the same curve.
In the same time interval,
the fundamental pulsation has almost disappeared
in the high energy range (100-700~keV), while it is still evident at 
medium and low energies.

\section{Spectral Analysis}

\subsection{Phase-resolved Spectroscopy}

Figure~\ref{grbm_panels} shows the high time resolution data from the
two experiments
in a series of panels, each displaying one stellar rotation cycle.
Figure~\ref{Ulysses_grbm_diff} shows the high resolution light curves in the
top panel, while the lower panel shows the ratio between the
curves (GRBM counts divided by \it Ulysses \rm counts).
The vertical dotted lines are separated by one spin period.
The BeppoSAX data declines rather smoothly during the first $\sim 40$
seconds, but even at early times, the modest dips are correlated
with the phase of the profound minimum which emerges after $\sim 40$ s.
The softer-spectrum {\it Ulysses} data shows no such correlation.
Indeed, the hardness ratio between the BeppoSAX and \it Ulysses \rm
counts (that is, between the flux in hard and soft X--rays; bottom panel),
seems to vary quite irregularly as a function of the spin phase
(that is related to the region of the neutron star) before $\sim 40$ s
and going towards a `see-saw' behaviour (see Feroci et al. 1999)
after that.

Although the subpulses clearly exhibit intrinsic variations over
successive cycles, we used the GRBM data in the last 8 panels
of Fig.~\ref{grbm_panels} ({\bf l} through {\bf s},
where the subpulse structure is clearly seen) to derive an average
folded pulse shape, shown in Fig.~\ref{folded_grbm}.
We preferred not to subtract any off-pulse continuum before folding 
our data. In fact, as can be noted in Figure 1 of Feroci et al. (1999),
the average flux during the last 8 complete cycles in our data
does not change very 
much, and we can therefore neglect the unbalance due to the first
cycles with the highest count rate. 
The centroids of the four sub-pulses are nearly evenly-spaced
by $\sim$1/5$^{th}$ of the spin phase, with a ``missing subpulse" at
the phase of the deep minimum.

We then used the real-time data to test for possible correlations 
between hardness and intensity within rotational cycles. 
In Figure~\ref{folded_ul}
(bottom panel) we show the (100--700~keV)/(25--150~keV) ratio 
folded on the 5.16-s rotation period for 
the portion of the light curve between $t = 50$ and 265 s 
(37385 to 37600 SOD), beginning shortly after the emergence of the
four-peaked pattern.  In the top panel of the same Figure we plot the 
folded \it Ulysses \rm light curve, over the same period of time. There 
is a clear anti-correlation between the two curves,
showing that the detected radiation is hardest at the intensity minima. 
We obtain similar results using the two BeppoSAX energy ranges 
(40--100 and 100--700~keV) and the same data as Fig.~\ref{folded_ul}
in shorter time intervals.
Furthermore, using the high energy data in the 
time period $t = 40$ -- 100 s, when the 
four-peaked emission pattern is most prominent, 
Feroci et al. (1999) found a 
`sawtooth' pattern in the hardness ratio between the two energy 
ranges available from the BeppoSAX GRBM  \hbox{[(100--700)/(40--100 keV)],}
featuring steep hard-to-soft
evolution across the deep minima in the light curve. 

Note that BeppoSAX and {\it Ulysses} measurements 
are not able to constrain spectral variations 
across individual peaks of the four-peaked pattern, because
of limited span of the high time resolution data set.  
In contrast, the Konus experiment recorded the hardness ratio 
\hbox{[(50--250)/(15--50 keV)]} with 256 ms time-resolution
out to $t = 230$ s (Mazets et al.~1999).  These data show a 
mild {\it positive} correlation of hardness with the intensity 
of the individual subpulses when the four-peaked pattern is most 
prominent ($t = 40$ to 100 s). 
After $t = 100$ s, the Konus light curves evolve toward
a flat-top/two-peaks pattern, with the initial four peaks 
``merging" earliest in the soft photon band (15 -- 50 keV).  
Although a general hard-to-soft trend is visible within the pulse since 
$t = 40$, the Konus hardness ratio becomes smoothly sawtooth-like 
for $t = 140$ -- 180 s, 
with rapid hard-to-soft evolution across the minimum.
Between $t =180$ s and 230 s, Konus data show a clear hardness-intensity
anti-correlation, resembling our Figure~\ref{folded_ul}.

\subsection{Time-resolved Spectroscopy}

The real-time data available from BeppoSAX and \it Ulysses \rm
and their serendipitous synchronization allow us to study the 
long-term spectral
evolution with a 1-s time resolution. In Figure~\ref{hr}
we show the hardness ratio between the GRBM (100--700~keV)
and the \it Ulysses \rm (25--150~keV, with a non negligible
contribution from energies between 10 and 25 keV) data.
These data were time-averaged over 5-s intervals in order to
show the general spectral trend, averaging over the spectral
evolution within each rotational cycle.
Interestingly, the spectrum gradually softens during the
first $\sim$40~s and then flattens when the large-amplitude
pulsations set in.
Finer details of the spectral variation in the early phase of the
event may be found in Hurley et al. (1999b), Feroci et al. (1999)
and Mazets et al. (1999).

As discussed above, both the GRBM and the \it Ulysses \rm  GRB detectors 
also provide energy spectra integrated over fixed time intervals,
with \it Ulysses \rm having shorter integration times than GRBM. 
In Table 1 we show the three shortest integration times which could
be used to construct joint spectra.  
Interval A covers roughly the first
$\sim 65$ s of the burst; Interval B, the next $\sim 130$ s;
and Interval C, the $\sim 130$ s after that, which lasts until the end of 
flare emissions. The  \it Ulysses \rm spectra were co-added
to obtain energy spectra integrated 
over approximately the same time intervals as the three GRBM energy spectra.
As specified in Feroci et al.~(1999), a 10\% systematic uncertainty
was added to the GRBM energy spectra, to account for the
uncertainty in the response matrix at such a large off-axis angle,
where the detector energy-dependent projected area is known 
with the above uncertainty.
Similarly, a 10\% systematic error was added to
\it Ulysses \rm spectra to account for differential non-linearity
in the multichannel analyzer.

This procedure allows us to extend the energy spectra presented in
Feroci et al.~(1999) down to about 20~keV, and to fit spectral
models over this extended energy range.
Interval A includes the initial hard spike. During this short event our
detectors were affected from pulse pile-up and 
deadtime effects, that are likely affecting the spectral form as well, and
that cannot be accounted by the systematic error added to the data. 
In addition, we know from the time-resolved spectral analysis 
(see Fig.~\ref{hr}) that a large spectral evolution is comprised in this
interval. For these reasons we don't present here spectral
results for interval A. From a technical point of view,
attempts to make a joint fit to the data of Interval A were unsuccessful,
whereas a reasonably good fit to the GRBM data alone was presented
in Feroci et al.~(1999) who employed optically-thin thermal 
bremmstrahlung (OTTB) plus power-law spectral functions, with a 
power-law photon index $\sim$1.5. 

For interval B\footnote{
We note that for this interval we don't expect the data to be
affected by pile-up effects. In fact, 
the GRBM electronics can easily handle a count rate  
larger than $\sim$4$\times$10$^{4}$ counts s$^{-1}$, whereas in this
interval the maximum 40-700~keV count rate is of the order
of $\sim$7$\times$10$^{3}$ counts s$^{-1}$.
In \it Ulysses \rm the microprocessor can only handle about 
$\sim$3$\times$10$^{3}$ counts s$^{-1}$ (the number actually depends 
strongly on the spectrum) so the electronics was definitely saturated 
as far as dead
time is concerned, and this was corrected for.  However, pulse pile-up
effects only seem to come in at much higher rates, as verified with
solar flares, so no correction for pile-up was required for interval B.  
}, we know that the spectral temperature probably 
varies between only slightly different values during each rotation cycle
(Fig.~\ref{hr}; \cite{feroci99}; \cite{mazets99}). 
We tested several spectral functions, and the fit parameters that we
obtain are presented in Table~2. From those values it appears that a
very satisfactory spectral fit cannot be obtained for the 
models tested: the minimum value for the
reduced $\chi^{2}$ is about 1.6 for 81 degrees of freedom. The minimum
$\chi^{2}$ spectral model is composed of two
blackbody (BB) laws, with temperatures of 9.3 and 20.3 keV, plus a
power law (PL) with photon index of 2.8.
This model gives a significantly better fit to the thermal component
of the spectra than does the traditional OTTB fitting function. 
In both models (two BB plus PL; and OTTB with $kT\simeq 23$ keV plus PL) 
the power--law accounts for approximately 10\% of the total energy 
above 25~keV.  \ In the case of two
blackbodies, the low-temperature component accounts for
about 85\% of the total energy above 25~keV.
 
Many fewer photons were detected during Interval C, 
thus the spectral fits in this interval are not as strongly constrained.
The models we tested and their best-fit parameters are shown in Table 3. \
The OTTB model cannot be reasonably excluded: according to an F-test,
the OTTB fit is an improvement over a cut-off power law, with 
$\sim$98\% confidence.

Note that the relative normalization between \it Ulysses \rm and GRBM was 
taken as a fitting parameter in intervals B and C.  This factor 
turned out to be rather insensitive to spectral models, and it had a 
very similar value in the two time intervals.

\section{Re-analysis of the SIGNE and ISEE-3 data on the March 5th event}

The 1979 March 5 event and the 1998 August 27 event 
had many similar properties, although they originated from different SGRs 
(e.g., \cite{hurley99b,mazets99}). 
In order to compare these two flares, 
we now reanalyze SIGNE data on the March 5 event, following some
of the same procedures that we applied to the August 27 event. 

The SIGNE gamma-ray experiments aboard the Prognoz and Venera spacecrafts
(\cite{barat81}) consisted of 64 cm$^2$ NaI(Tl) detectors operating in 
various energy ranges between $\sim$ 50 keV and 1 MeV.
For our purposes, the best SIGNE spectral data for the March 5 event
consist of Venera 130--205 and 205--353~keV count rates for the
first 32~s from the event onset.
In Fig.~\ref{signelc} the light curve in two energy ranges 
(130--205 and 130--353~keV) is presented with 250~ms time resolution. 
In Fig.~\ref{venera_march5} we show a 1 s resolution SIGNE
light curve and hardness ratio (203--353~keV counts divided by
130--203~keV counts) folded with the $\sim$8.0~s (\cite{barat79})
period. Due to onboard compression of the number of counts, errors were
derived for these lightcurves by means of Monte Carlo simulations.

If we compare Fig.~\ref{venera_march5} for the March 5 event and
Fig.~\ref{folded_ul} for the August 27 event, we note the same
type of spectral evolution, namely harder spectra during the
interpulse than during the pulse.  This result may be contrasted with
that of Mazets et al.~(1982), who fitted an OTTB function to the energy
spectra of the pulsating component of the March 5 burst. 
They found the interpulse
spectra to be slightly \it softer, \rm by 4 keV, than the pulse spectra.
However, Mazets et al.~(1982) fit their energy spectra over the
30--300 keV range, and had only 4 s resolution.

Given the excellent fit of a trapped fireball to the August 27
lightcurve (\S \ref{envel}) let us compare this model, and a simple 
exponential decay model, with the March 5 lightcurve.  
In order to do this we need a longer March 5 lightcurve. At this scope
we used the ISEE-3 lightcurve (for energies above 50 keV)
published by Cline et al. (1982).
Figure \ref{fire_venera} 
shows the ISEE-3 data (Fig.~\ref{fire_venera}b shows a lightcurve with
binsize close to the 8-s oscillation period on a linear timescale), 
with superposed two exponential ($\tau$=60 and 80~s) 
and the best-fit trapped fireball parameterization:
[$L_{X}(0)$=(350$\pm$6) counts/s, $t_{evap}$=(163$\pm$5) s
and $a = 0.71 \pm 0.01$]. In doing this fit we started soon after 
the initial spike, that is as soon as the high amplitude oscillations
set in, consistently with what we did for the giant flare from SGR~1900+14. 
Although the decay can be followed only over a
factor $\sim 30$ in flux, the light curve does appear to break more sharply
than can be fit with a simple exponential. Note also that an intermediate 
smooth tail (lacking large-amplitude oscillations) is much briefer in the 
March 5 event than it is in the August 27 event.

\section{Discussion}

The August 27 giant outburst exhibited three phases:  
a short and very intense initial spike lasting $\sim 0.5$ s;  a
softer-spectrum, smoothly declining  tail lasting $\sim 40$ s; and
a phase of large-amplitude oscillations at the 5.16 spin period of the
source. We now briefly outline how these phases can be interpreted in
the context of the magnetar model. 
The initial spike appears to have involved the free
expansion of relativistic pair plasma -- a miniature version
of the gamma-ray fireballs that are observed at cosmological distances.
After $\sim 40$ s, the envelope of the light curve asymptotes
to a form that is consistent with a cooling $e^\pm$ fireball,
trapped by the closed magnetic field lines of the neutron star. 
The excellent fit to the light curve gives strong evidence that most of the
burst energy was either blown away from the star or  
deposited in the magnetosphere during the initial second.  However, the
intermediate 40-s tail is suggestive of an extended pair atmosphere, which 
requires continuing energy output from the star.  Finally, we
ascribe the emergence of the striking four-peaked pattern to
a transition from a pair-dominated to an ion-dominated photosphere.  
Collimation of the X-ray jets is a direct consequence
of the cooling of a trapped fireball in a superstrong magnetic
field (\S 6 in TD95).  The repetition of the four-peaked pattern over 
the star's 5.16-s rotation period points to a strong multipolar component
in the magnetic field of SGR 1900+14.  We now discuss each phase in
more detail, leaving a more extensive treatment
of the theoretical issues to Paper II.

\subsection{The precursor}

Panel {\bf a} of Fig.~\ref{grbm_panels} shows the short
precursor observed $\sim$0.4~s before the event by the \it Ulysses \rm
detector.  The precursor is barely detected in the BeppoSAX data, which
implies a soft spectrum, probably resembling the much more common
short bursts from this source.  It was conceivably the SGR
analog of the ``foreshocks" which often precede large earthquakes
[e.g., \S 7.2 in Sholz 1990].  
Just three months prior to the giant
flare, the source emerged from a quiescent period lasting several years,
and started to emit short, recurrent soft bursts. If the origin of this
activity is localized crustal adjustments, compensating for magnetic
stresses, the precursor could have been the last (failed) attempt
to compensate for the extensive stress which caused the
giant flare, a few hundred milliseconds later.

\subsection{The initial hard pulse}

The hard spike of the August 27 flare had a 
full width at half maximum duration of $\sim 0.3$~s, 
as measured by the \it Ulysses \rm lightcurve 
(Fig.~\ref{grbm_panels} and \ref{Ulysses_grbm_diff};
see also \cite{hurley99a}; \cite{mazets99}).  Its peak luminosity
exceeded $4\times 10^{44}$ erg s$^{-1}$ at a distance of 10 kpc (Mazets
et al. 1999). This luminosity is intermediate 
between an Eddington-limited thermonuclear X-ray flash on an
accreting neutron star, and a bright gamma-ray burst observed at 
high redshift. 

The short duration, high luminosity and hard spectrum of the initial spike
indicate that a {\it relativistic} outflow was driven from the star.  
To see this, consider an expanding plasma that contains comparable 
energy in radiation and in the rest energy of (baryonic) matter: 
$E_\gamma\approx Mc^2$.  \ When the fireball expands to radius $R$, 
it has Compton optical thickness 
$\tau\sim \sigma_T M/ (4 \pi R^2 m_p)$.  The observable duration of the
photon emission is $\Delta t = R_o/c$, where $R_o$ is the radius at 
which $\tau\sim 1$:
\begin{equation}\label{delt}
\Delta t \sim {1\over c^2} \, 
\left({\sigma_T \, E_\gamma \over 4\pi m_p}\right)^{1/2}
\sim 2.0\,\left({E_\gamma\over 10^{44}~{\rm
erg}}\right)^{1/2}\;\;\;\;\;\;{\rm s}.
\end{equation}
This is a several times longer than the observed duration of the
August 27 event hard spike.  Moreover,
eqn.~(\ref{delt}) {\it underestimates} $\Delta t$, because 
it neglects the opacity of pairs that may be present, and because
adiabatic expansion converts photon energy into baryon kinetic energy,
reducing the radiative efficiency 
$\varepsilon_{rad} \equiv (E_\gamma/M c^2)$, and thus increasing $\Delta t$ 
by a factor $\varepsilon_{rad}^{-1/2}$.  We conclude that this kind
of marginally-relativistic fireball cannot match observations: 
the initial spike of 
the August 27 flare must have been
powered by a very clean energy source:  $M c^2 < E_\gamma$.
Independent evidence for relativistic particle ejection comes from
the observed synchrotron afterglow (Frail, Kulkarni \& Bloom 1999).  

 Rapid, high-amplitude modulations in the flux of very hard
photons ($E_\gamma > 250$ keV) were observed from 0.2 to 0.6 sec 
after the burst trigger (Mazets et al. 1999), with 
peaks narrower than $\sim 0.01$ s.  This fine time-structure constrains 
the baryon loading even more severely. When taken together
with the hard spectra, this suggests that the light curve may directly
trace the release of energy by the neutron star, with 
discrete 0.01-s ejections of $e^\pm$-dominated plasma.
Similar rapid variability was observed near the peak 
of the March 5 event (Barat et al.~1983).

Giant SGR flares are thought to occur when the evolving stellar
magnetic field reaches a point of instability and catastrophically 
shifts to a lower-energy state. A magnetic field 
$B \ga B_{frac} = (4\pi\theta_{max}\mu)^{1/2} \sim
2\times 10^{14}\, (\theta_{max}/10^{-3})^{1/2}$ G can induce 
a shear deformation along equipotential surfaces up to the
critical strain angle $\theta_{max}$ at which the crustal lattice
fractures (TD95).  
However, such a field is far too weak to induce energetically
significant compressional or vertical displacements in the deep crust. 
The time structure of the initial spike indicates that the release
of energy is limited by frictional and inertial forces:  a fireball
resulting from a sudden readjustment of the external field such as an 
exterior reconnection would last only $\sim R_{NS}/c \sim 10^{-4}$ s 
or slightly longer (TD95).  In other words, the giant outbursts
are probably driven by internal (rather than external) magnetic
stresses.  A $10^{15}$ G magnetic field can move
the core material at a speed $\la B/\sqrt{4\pi\rho}$ (TD95), and so the
total duration of the spike is greater than
\begin{equation}
\label{spiket}
t_{\rm spike} \ga {R (4 \pi \rho)^{1/2} \over B} = 0.1 \, R_{10}  \,
\rho_{15}^{1/2} \, B_{15}^{-1}\;\;\;\;\;\;{\rm s}.
\end{equation}
Here, we scale to a stellar radius  10 $R_{10}$ km, an interior density
$10^{15} \, \rho_{15}$ gm cm$^{-3}$ and an interior field $10^{15} \,
B_{15}$ Gauss.  This implies a lower bound $B > 2\times 10^{14}$ G to the 
internal magnetic field in SGR 1900+14. 

\subsection{The Smooth Decay}\label{smooth}

The August 27 lightcurve, averaged over the pulse period,
decreased nearly monotonically after the initial spike
(Fig.~\ref{smooth_decay}).  
The measured flux significantly exceeds the backward extrapolation 
of the trapped fireball lightcurve during the first $\sim 40$ s
(Fig.~\ref{fire_3decays}).

The fraction of the post-spike fluence carried by this excess component
can be calculated as follows.
A good fireball fit in the waveband  40--100 keV is given by 
$t_{evap} \soe 410$ s, $a \soe 0.78$, and
$L_X(0) \soe 1.1 \times 10^{-5}$ erg cm$^{-2}$ s$^{-1}$. This implies
a total fireball fluence in this waveband of 
$(1-a)\,L_X(0)\,t_{evap} \soe 1 \times 10^{-3}$ erg cm$^{-2}$.  The 
fireball energy for isotropic emission is then 
$\sim 1.2\times 10^{43}\, D_{10}^2$ erg, assuming that SGR 1900+14 is 
$10 \, D_{10}$ kpc from Earth.  (Substantial energy is also emitted 
outside this waveband.)   In comparison, the excess emission 
from 3--40 seconds post-trigger carries a fluence 
$\sim 2.3 \times 10^{-4}$ erg cm$^{-2}$, 
constituting $\sim 20$\% of the net output of 40--100 keV photons
after 3 s.  

The light curve shows only a mild rotational modulation during this
40-s smooth tail (Figs.~\ref{fig_psd}, \ref{grbm_panels}
and \ref{Ulysses_grbm_diff}). The onset of 
large amplitude pulsations coincides with the decay of the excess
emission (Fig.~\ref{fire_3decays}).  The spectrum was significantly 
{\it harder} during the first $\sim 40$ s than during
the remainder of the burst, when the temperature declined more
slowly (Fig.~\ref{hr} and Mazets et al. 1999).  
These observations suggest that 
the smooth excess was powered by ongoing Alfv\'en 
heating-in a pair-dominated corona surrounding the
trapped fireball.  The emergence of the pulsations points to
a contraction from an extended, pair-loaded photosphere to one
dominated by electrons and ions.  The radiative flux out of the
trapped fireball is collimated along open bundles of magnetic flux
(\S \ref{fourpeaksec}).  During this final cooling phase, the 
temperature of the emergent radiation becomes strongly regulated, 
for two reasons:  the trapped fireball
cools at a rate proportional to its surface area; and photon splitting
becomes ineffective below a critical temperature of $\sim 11$ keV (TD95).

Given a strong enough outflux of Alfv\'en waves, a non-thermal pair atmosphere 
beyond the trapped fireball is inevitable. Interacting Alfv\'en 
modes undergo a turbulent cascade, which forces the 
current density slightly above the value that can be supported 
by the available pairs, driving electromotive heating of 
pairs (Thompson \& Blaes 1998).  
The pairs also cool via Compton scattering off photons emerging from
the trapped fireball below, and equilibrate to a temperature only slightly
higher. The Compton parameter
${1\over 4}y = \tau_{es}^2\,k(T_{e^+}-T_\gamma)/m_ec^2$ is 
determined self-consistently by balancing these heating and cooling rates.
The pair temperature must exceed $kT_{e^+} \sim 20$ keV in $B \sim 
10 \ B_{QED}$ so that the Comptonized spectrum self-consistently
generates an optical depth exceeding unity in pairs, via photon collisions, 
$\gamma + \gamma \rightarrow e^+ + e^-$ (cf.~Fig. 4 in TD95; Paper II).
After the cascade power $L_{cas}$ drops below the luminosity of the
cooling fireball, this pair atmosphere evaporates and the scattering 
photosphere contracts to the outer
boundary of the trapped fireball, where the opacity is dominated by
ion-electron plasma (TD95).  

We conclude that the decreasing excess flux during the 40-s smooth tail 
provides evidence for a late (post-spike, $t> 1$ s) seismic excitation 
of the neutron star.  
Such active internal modes can couple strongly to
exterior Alfv\'en waves (Blaes et al. 1989).
Magnetospheric Alfv\'en 
excitation at a characteristic radius exceeding $R_{max}$ implies 
an internal mode period larger than $P \sim 3R_{max}/c \sim 1\,
(R_{max}/100~{\rm km})$ msec.   A variety of
non-radial neutron star modes have such low frequencies (e.g., McDermott,
Van Horn \& Hansen 1988; Reisenegger \& Goldreich 1992) including     
trapped Alfv\'en modes in the liquid interior of a magnetar 
(cf.~eq.~[\ref{spiket}]).  
Toroidal modes in the crust are especially likely to be excited by the
large propagating fractures of giant flares;
and since they are confined to the thin crust, they
can readily Alfv\'en-damp (Duncan 1998).
This seismic energy could have been deposited at the initial
outburst (during the hard spike); or, alternatively, it involved
a continuing crustal motion at the fracture site (with the rate of
dissipation of Alfv\'en energy in the magnetosphere being smoothed
over the collision time between waves).

\subsubsection{The Four-Peaked Repetitive Pattern}\label{fourpeaksec}

The clearing away of the wide-ranging, optically-thick magnetospheric
plasma, and
the opening of the first optically-thin channels extending down to near
the star's surface is dramatically shown by both {\it Ulysses} and
BeppoSAX data at 35 to 50 seconds after the burst onset
(Figs.~\ref{grbm_panels} and \ref{Ulysses_grbm_diff}).
The subsequent X-ray light curve shows a strong, four-peaked
modulation on the 5.16-s rotation period.
The peaks are almost evenly-spaced at 1.0 s
intervals (Figs.~\ref{grbm_panels} and \ref{folded_grbm}),
giving rise to an strong fifth harmonic near
1.0 Hz in the power spectrum (Fig.~\ref{fig_psd} and \cite{feroci99}).

As can be seen in panels {\bf h} through {\bf k} of
Fig.~\ref{grbm_panels},
the strongest peak, peak 2, emerges first; then peaks 1, 4 and 3 appear
successively during each of the following three rotation cycles.
 Some evidence for the broad trough that separates peaks 1 and 4 appears
much earlier,
 (see panels {\bf b}--{\bf e} of Fig.~\ref{grbm_panels} and
\cite{mazets99}). 
This suggests that the magnetosphere has a gross dipole imbalance, with
more
entrained energy on the side which faces {\it away} from Earth during
the trough.

The narrow peaks and dips extend significantly above and below
the extrapolation of the earlier, less modulated light curve
(Fig.~\ref{Ulysses_grbm_diff} and Fig.~1 in Feroci et al. 1999).  
In some places, the flux varies by
almost an order of magnitude peak to dip.
These features can accurately be described as collimated X-ray jets,
which are swept past the line of sight as the neutron star rotates. 
The rotationally-averaged X-ray flux shows no dramatic change during
the emergence of the pulsations (Fig. \ref{smooth_decay}),
which implies that the the pulsations result from a
redistribution of a (nearly) conserved X-ray luminosity.  This behavior is
consistent with our inference that 
the thermal X-rays escaping from the trapped fireball are 
reprocessed through an extended pair-dominated halo during the 
first 40 s.

What causes these X-ray jets to form?  Their phase stability suggests
that they are
tied to surface features on the neutron star.  We now argue that
the burst light curve provides a template of the neutron
star's surface magnetic field, and indicates 
that it has a complicated, multipolar structure.

This time structure is almost certainly {\it not} driven
by a 1-second oscillation of the neutron  star.  The pattern
repeats coherently with the 5.16-s rotation period of the source:
four peaks per cycle, with the second one strongest,
and a ``missing peak" during the main trough.  Furthermore
Fig.~\ref{folded_grbm} shows that the peaks may be
not precisely spaced at even intervals; and the fourth peak is ``split"
by about 0.3 s.  
The striking 1-to-5 relationship between the peak spacing and
the 5.16-s spin period might suggest some relation between the two.  
However, we are not aware of any mechanism that could produce such an
effect, and we are led to the conclusion that it is probably 
a coincidence.  This conclusion is supported by the
Konus observation of the late time evolution of the four-peaked
structure (\cite{mazets99}), in which there is a convergence to a 
double-peaked structure.  Plausible internal modes of the neutron
star have periods much shorter than 1.0 s. \ 
The longest-period mode in the crust is the fundamental toroidal 
oscillation with $P(_2t_0)\sim 0.03$ s (Duncan 1998);  and 
the lowest-frequency Alfv\'en modes in the liquid interior
are also probably too fast [eq.~(\ref{spiket})]  (Note that interior fields
much stronger than the exterior dipole field are expected to result from
post-collapse $\alpha$--$\Omega$ dynamos; DT92, TD93.).

Independent evidence for a multipolar field structure at the surface
of SGR 1900+14 comes from its quiescent X-ray light curve.
Woods et al.~(1999b) found a four-peaked light curve in 1997 May
that resembles Fig.~\ref{folded_grbm}, except that the strongest peak
is peak 1 (just after the minimum) rather than peak 2. 
A magnetar's light curve during active periods could be dominated by
persistent magnetospheric currents, driven by twisting motions of the
crust (\S 5 in Thompson et al. 1999).  This mechanism potentially explains 
the X-ray brightening and dramatic light-curve change
in the aftermath of the August 27th flare (Woods et al.~1999b;
Murakami et al. 1999).
Because the crust is nearly incompressible and stably-stratified,
with degeneracy pressure in hydrostatic balance with gravity that is
much
stronger than any lattice shear or magnetic stresses, a large-scale
propagating
fracture of the neutron star crust must involve a predominantly
rotational motion.

Collimation of the X-ray flux from a trapped fireball is discussed
in Paper II.  We summarize the relevant physics here.  
The two photon eigenmodes are linearly polarized when vacuum
polarization dominates the dielectric properties of the medium.
The Compton scattering cross section of the extraordinary mode (or E-mode:
$\delta{\bf E}\cdot{\bf B}_0 = 0$) is strongly suppressed:
$\sigma_E = (\omega m_e c /eB_0)^2\,\sigma_T$ (e.g. Herold 1979).
This suppression greatly increases the radiative transport rate
both from the surface of the neutron star (Paczy\'nski 1992) and
across a confining magnetic field (TD95).
However, the scattering opacity of the E-mode rises rapidly with radius
(e.g., $\sigma_E \propto B^{-2} \propto R^6$ in a dipole geometry).
As a result, most of the radiative transport occurs close to the surface
of the star.  The rapid growth of the E-mode opacity then provides
a mechanism for self-collimation:  the E-radiation can escape only by
pushing the suspended matter to the side.  A second effect 
involves the conversion of the E-mode to the ordinary (or O-mode;
$\delta{\bf B}\cdot{\bf B}_0 = 0$), which occurs effectively via 
both photon splitting and Compton scattering near the E-mode photosphere  
(TD95; Miller 1995).  Because O-mode scattering is not 
magnetically-suppressed, the energy flux converted into the O-mode
is tremendously super-Eddington, and it must drive a diffuse 
hydrodynamical jet, outward along channels of magnetic flux that are 
optically-thin to the E-mode. 
An estimate of $\dot{M}$ is given by the condition
that the relativistic, O-mode driven wind has Thomson optical depth near
unity across a height $\sim R_\star$.  This implies
$\dot M c^2 \sim (GM_\star /R_\star c^2)^{-1} \, (\Omega_j/4 \pi) \, L_{edd}$,
where $\Omega_j$ is the outflow or jet solid angle
(cf.~\S 6.4 of TD95).

The jet sub-structure is more pronounced, 
and more quickly emergent, in the light curves at higher energies 
(e.g. around 37370-37390 SOD in Fig.~\ref{Ulysses_grbm_diff} and 
at $t\sim 40$ s in Fig.~7 of Mazets et al.~1999).  This
is consistent with advection of thermal X-ray photons from hotter
regions, closer to the surface of the neutron star.

\subsection{Origin of the Hard Spectral Component}

The combined  \it Ulysses \rm and GRBM lightcurves provide clear
evidence for a hard component in the spectrum of the August 27 event.  
The early appearance of modest dips, at a phase coincident with the late 
deep minima, only in the GRBM data (hard X-rays) points to a site
closer to the star for the \it early \rm hard emission than for the 
soft one.    
In addition, the best fits to the time averaged spectra (see \S 4.2)
confirm the persistence of a non-thermal component at times $\goe$70~s
after the event onset.  This hard emission, which occurs
well after the initial spike, may be related to the persistent
Alfv\'en wave dissipation that (we have argued) powers the excess 
emission during the smooth 40-s tail (\S \ref{smooth}).  However, the
persistence of the hard component is more consistent with particle
acceleration driven by the hyper-Eddington photon flux:  for example,
at the boundaries of the X-ray jets, far out in the magnetosphere where
the photon pressure begins to exceed the magnetic dipole pressure.
Further evidence for 
radially extended hard emission is provided by the time-resolved hardness
ratio: the spectrum is hardest during the broad troughs that occur at 5.16-s
intervals (Fig.~\ref{folded_ul}).

A hard spectral component (photon index $\simeq 1.5$) was previously
reported in the first part of the outburst (Feroci et al.~1999).  
This hard spectral index is consistent
with passive cooling of very energetic charges.  Outside a radius
$R \sim 70\,R_{NS}\,(B_{dipole}/10\,B_{QED})^{1/2}\,
(L_X/10^{44}~{\rm erg~s^{-1}})^{-1/4}$, the energy density in the
escaping X-ray photons exceeds the energy density of the dipole
magnetic field.  This far from the neutron star, energetic charges cool
primarily by inverse Compton scattering.   Closer to the star, 
synchrotron emission can contribute significantly to cooling.  Observations of
future outbursts at energies $\gg m_ec^2$ will be sensitive to the 
compactness of the non-thermal emission region, and thus will be able 
to distinguish between these two regimes.

\subsection{Comparisons with the 5 March 1979 Event}

There are strikingly similarities between the giant SGR outbursts
observed on 1979 March 5 and 1998 August 27. \
Both events radiated $\sim 10^{44}-10^{45}$ erg in X-rays and gamma rays,
which exceeds the measured fluences of all other SGR outbursts 
by $\gtrsim 2$ orders of magnitude.\footnote{The second 
longest of the repeat bursts from SGR 0526-66 (duration 1.5 s) occured 
the day after the March 5 event (Golenetskii et al. 1987).  This burst 
could have been an aftershock on the same fault responsible for the
giant outburst.}
Both events began with a brief and very intense gamma-ray spike, which
was followed by a softer tail more than a thousand times longer in
duration than ordinary SGR bursts.  Also, in both events, the bursting flux 
was modulated by a large amplitude at the rotation period of the star
(with two prominent sub-pulses in the March 5 event as compared with
four in August 27), with similar intensity-hardness anticorrelation
in the main pulses.
Considering the many other characteristics shared by these  
two sources, there is compelling evidence that the two SGRs are physically 
similar stars which evolved through similar stages to reach  
similar catastrophic instabilities.   Moreover, the discontinuous 
distribution of outburst energies/durations suggests that the giant flare
instability is {\it qualitatively distinct} from the instability 
of ordinary SGR outbursts; i.e., different in kind, not 
just in degree.  It is not known how this sort of bimodal behavior
could be driven by an external trigger like accretion; but promising
mechanisms are provided in the magnetar model (Paper II). 
 
The most striking {\it difference} between the August 27 and March 5
events can be seen by comparing the light curves of Figure 
\ref{Ulysses_grbm_diff} with the 50--150 keV March 5th light curves 
recorded by the Konus experiment, onboard the Venera 11 and Venera 12 
spacecraft (Figure 2 in Mazets et al.~1979, hereafter ``M79",
and our Fig.~\ref{signelc}).   Almost 
immediately after the hard spike, the March 5th light curve 
is deeply modulated on the rotation period of the star, with a smooth, 
hump-like pattern of pulse and interpulse.  The interpulse is in phase 
with the initial hard spike, or nearly so. 
Over eight succeeding rotation cycles, the peak of the main pulse 
diminishes more steeply with time than does the peak of the interpulse. 
This contrasts with the smooth, only slightly rotation-modulated, 
quasi-monotonic flux decline seen 
during the first $\sim 40$ s of the August 27th event, followed by 
emergence of the four, strongly-peaked X-ray ``jets".  

If strong, jet-like modulations on
$\lessim 1$s time scales emerged in the March 5th light curve {\it after}
$\sim 30$ s, as they did in August 27, then they would not have been
detected: the time resolution of the M79 light curve
is not good enough.  At earlier times, the M79 data argue against any
such fine substructure.  But there is a suggestion of 1-s X-ray jet emergence
as early as $t\sim 10$ s in the higher-energy (130 - 205 keV)
data acquired by the French SIGNE experiment, also aboard Venera 12
(Fig.~\ref{signelc} and Barat et al.~1979, hereafter ``B79").
According to B79, ``four main
peaks exist [for the first four rotation cycles], each of which appears
to have a substructure of double or triple peaks of about equal intensity,
spaced $\sim 1$ s apart."   Figure 2 of B79, which is a smoothed version
of the lower panel of Fig.~\ref{signelc}, suggests that the main pulse
splits into three jets, and the interpulse into two.

A very strong hardness-intensity correlation within the jets
would be necessary for the suggested 1-s jets of B79 to be
reconciled with the smooth light curves of M79 at lower energies.
Thus jetlike substructure, if real, may be more profound in data sets
including higher-energy photons.  We searched for this effect.
The top panel of Fig.~\ref{signelc} shows the SIGNE light curve for
photon energies ranging up to 353 keV, based upon unpublished, archival
data.
We folded this light curve on the 8.0 s rotation period of the
star, including just the last three rotation cycles, since there is a
suggestion that the substructure emerges only $\sim 10$ s after the
event's onset; cf.~B79 Fig.~2. (When folded over all four observed cycles,
the pattern is even smoother.)
We conclude that a standard folding procedure applied to the SIGNE data
provides no compelling evidence for 1-s substructures in the March 5
event.

All the observations concur that any smoothly-declining, unmodulated component 
in the March 5 tail lasted for $\lessim 10$ s (as opposed
to $\sim 40$ s for August 27) and involved a portion 
of the post-spike emissions that is much smaller than the 
fraction 0.2 estimated for the August 27 event. 

One final note on the rates of giant flares.
The observed number of events (2 giant flares out of 4-5 known SGRs)
is consistent with the hypothesis that flares 
occur in {\it all} known SGRs at a rate $\sim 1$ per 20 yrs, per source. 
But we cannot rule out the possibility that there exists a sub-population 
of SGRs (or of non-SGR magnetars) which never experience such outbursts.

\section{Summary and conclusions}

In this paper we present an analysis of the
data on the giant flare from SGR 1900+14 available from the
BeppoSAX Gamma Ray Burst Monitor and from the \it Ulysses \rm
Gamma Ray Burst detector.
We observe three clearly distinct stages in the August 27 event.
During the first 0.5 seconds the star emitted {\it an intense spike
of hard gamma-rays} which saturated all detectors.  
The  fluence in this hard spike was at least half the total photon fluence
of the event at energies above 15~keV (Mazets et al. 1999).  
Its duration and fast ($\lte 4$ ms) rise time 
are consistent with a magnetically-driven instability 
in an ultramagnetized ($B > 10^{14}$ Gauss)  neutron star,  
involving a large propagating fracture in the crust.  
 After the hard spike, the flare's light curve shows two distinct phases:
a {\it smooth, soft tail} that lasted $\sim 40$ s, followed by
the emergence of a striking pattern of {\it four large-amplitude peaks}
that repeat coherently on the 5.16-s rotation period of the neutron 
star.  During both these phases, the rotation-averaged emissions declined 
nearly monotonically, and released a total energy not much larger than
the initial spike (cf. Mazets et al.~1999).  Taken together, these
observations suggest that most of the burst energy was released during 
this initial half-second.  
Half or more of the energy escaped promptly 
in a {\it relativistic} outflow of electron-positron pairs and hard gamma
rays, and the remainder was trapped in the star or its vicinity (TD95).
Independent evidence for a particle outflow comes from the detection
of a radio afterglow which faded in the days following the event  
(Frail, Kulkarni \& Bloom 1999). 

     The rapid drop in X-ray flux at the end of the burst
gives strong evidence for an emitting resevoir of energy that 
evaporates completely in a finite time.   The
envelope of the light curve, from 40 s to the termination of the
burst, is well fit by the contracting surface of a thermal photon-pair plasma,
confined by the strong magnetic field
and anchored to the rotating surface of the star (eqn.~(1); cf.~TD95).
About 80\%  of the post-spike fluence is associated with this phase of 
large amplitude pulsations.  
    
    The smooth, soft tail has a harder spectrum than the subsequent emissions,
and the measured X-ray flux lies well above the trapped fireball light curve
that fits so well after $\sim$ 40 s.  The flattening of the light curve, 
and the decay of this excess emission, occurs simultaneously with the 
emergence of the four large-amplitude sub-peaks.  
Given these facts, we associate the excess emission
(representing $\sim 20$\% of the post-spike fluence) with
Compton heating by an extended corona of hot pair plasma (Paper II).
This heating must occur outside the trapped fireball, and is plausibly
driven by a persistent output of Alfv\'en waves from the neutron star.
In this model, the appearance of the large-amplitude pulsations 
coincides with the evaporation of the pair plasma and a contraction
of the scattering photosphere to the surface of the trapped fireball,
which sits congruent with the confining magnetic field.   The
persistent Alfv\'enic heating could be driven by crustal shear waves,
excited at the initial instability (Blaes et al 1989; Duncan 1998);  
or by persistent 
but declining seismic activity in an active fault zone.

    The large-amplitude pulsations require narrow collimation of
the X-ray flux, which is provided by the rapid increase in the E-mode
scattering cross section with distance from the neutron star.  
In addition, the O-mode  has a much stronger coupling to matter than the
E-mode and will flow hydrodynamically, at mildly relativistic
speed, along open channels of magnetic flux.  Both modes are emitted
through X-ray fan jets, which are
fed from below by rapid radiative diffusion of E-mode photons out
of the trapped reservoir of hot plasma (followed by conversion to the O-mode
through scattering or splitting).  Collimated
X-ray emission is a natural consequence of super-Eddington 
radiative transfer in magnetic fields stronger than 
$B_{QED}= 4.4 \times 10^{13}$ Gauss (\S 6.4 in TD95; \S \ref{fourpeaksec};
Paper II). 

    The light curve of the August 27 flare provides valuable information
about the magnetic geometry in SGR 1900+14.  The four sub-pulses
require the presence of higher magnetic multipoles -- perhaps the most
tangible evidence yet for a complex magnetic geometry in an isolated
neutron star.  In this picture, the observation of two (rather than four)
sub-pulses in the March 5 event simply reflects the chance location of the
outburst, and the differing structure of the magnetic field in that
second source.

    The joint analysis of the \it Ulysses \rm and GRBM spectra 
shows that a hard spectral component persisted long after the initial
hard spike.  This conclusion is 
corroborated by the time-resolved hardness ratio, which shows that
the burst spectrum is hardest during rotation-cycle minima.
We have noted two possible mechanisms for accelerating non-thermal
particles:
persistent Alfv\'enic heating, for which there is a clear evidence at 
times $\loe$40~s; or magnetohydrodynamic turbulence at the boundaries 
of the X-ray jets responsible for the four-peaked repetitive pattern.

Finally, we re-analyzed the SIGNE data of the first 32~s of the 
1979 March 5th event from SGR 0526-66, and ISEE-3 data for the first
$\sim$150~s of the same event. We found more properties that 
are shared with the August 27 event, lending support to the
hypothesis that similar physical instabilities drove both outbursts. 
In particular, we find that the spectral evolution 
across the pulse is strikingly similar to that of the giant flare 
from SGR 1900+14.  A trapped fireball model appears to fit 
the March 5 light curve better than a simple exponential, just as 
in the case of the August 27 event, although the short duration of 
the available light curve and the small dynamic range ($\sim 30$) in flux 
do not allow a definitive conclusion.
We also considered the suggestion of jetlike substructure on 1-s 
timescales in the SIGNE March 5th light curve (Barat et al.~1979).  
By a standard epoch-folding approach, we found no statistically-compelling 
evidence that such substructure emerged during
the first 32 seconds of the event.

We conclude that the magnetar model offers a promising framework for
understanding giant flares from SGRs.  The basic picture of an initial
relativistic outflow followed by diffusive cooling of a trapped, thermal 
fireball seems robust.  But a detailed understanding of the radiative transport
will involve mastering several complications, including: the effects of bulk 
hydrodynamical streaming, the large variations of the radiative 
transport coefficients arising from inhomogeneities in the background
magnetic field, a non-dipolar magnetic geometry, and the behavior of 
a pair atmosphere that is simultaneously heated by Alfv\'en waves and 
irradiated with X-rays from a trapped fireball.  Much work will be needed 
to test the physical mechanisms suggested here and in Paper II, and
to make a quantitative comparison with alternative models.

\acknowledgments
The authors thank the anonymous referee for suggestions 
improving the paper.
The authors are also grateful to C. Barat for the use of SIGNE data.
MF acknowledges useful discussions with F. Frontera, E. Costa, M. Tavani, 
E. Massaro, G.L. Israel and thank L. Amati for his unique support 
in the GRBM data reduction and analysis.
MF is supported by the Italian Space Agency (ASI).
KH is grateful to JPL for \it Ulysses \rm support under Contract 958056,
and to the NASA Astrophysics Data Program for supporting the integration
of BeppoSAX into the IPN under NAG5-7766. 
RD acknowleges support from Texas Advanced Research Project
grant no.~ARP-028 and NASA grant NAG5-8381.  CT is supported by
NASA grant NAG5-3100, and the Alfred P. Sloan Foundation.
BeppoSAX is a program of the Italian Space Agency (ASI), 
with participation of NIVR, the Dutch Space Agency.

\newpage

\centerline{FIGURE CAPTIONS}

FIGURE 1:
Background-subtracted \it Ulysses \rm  and BeppoSAX light curves
of the decay portion of the 1998 August 27 giant flare, showing the
overall patterns of decay in the three energy bands. The red continuous
lines indicate exponential laws with the specified time constants,
$\tau$. Blue, dashed lines indicate exponential laws with exchanged
time constants (70~s where $\tau$ is 78~s, and vice versa).

\begin{figure}
\figurenum{1}
\plotone{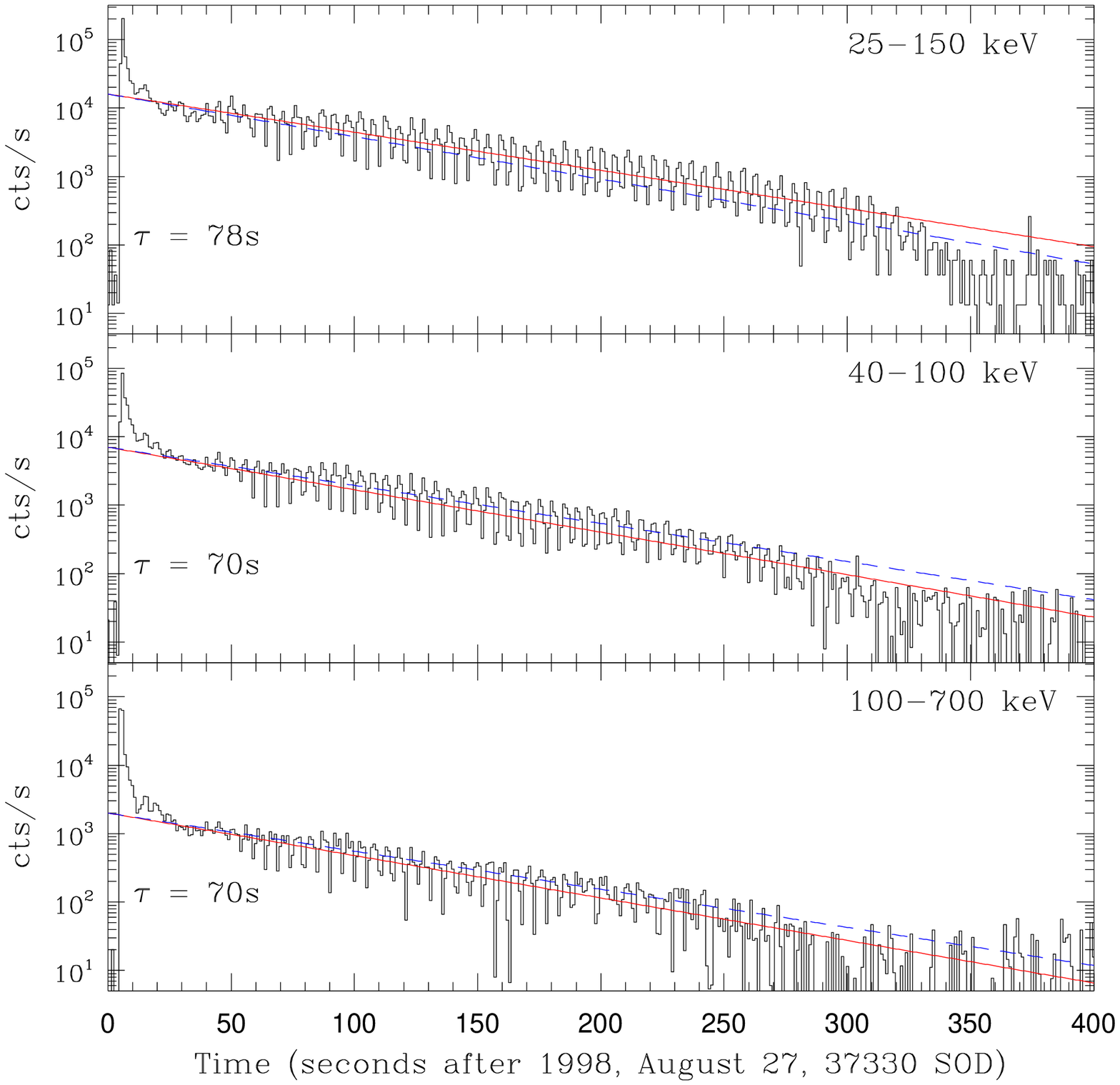}
\caption{
}
\label{3decays}
\end{figure}

FIGURE 2:
Analytical model of a cooling trapped fireball (eq. [\ref{lxt}])
superimposed on the \it Ulysses \rm light curve,
background subtracted and binned to 5-s intervals,
the closest available approximation
to the star's spin period. Panel (a) shows the complete light curve
while Panel (b) shows a zoom-in of the final drop.
The green dashed line shows the result of a
simultaneous fit (see text) made to the evaporation time
$t_{evap}$, the fireball index $a$ and the fireball luminosity
$L_{X}(0)$ for the time interval from 50 to 450~s after the peak
(that provides a  value too large for $t_{evap}$). 
The blue dashed line shows the result of a similar fit but to the
time interval  from  250 to 450~s after the peak.
The red curve shows the best fit to the curve in the time interval from 
50 to 450~s, constraining $t_{evap}$ in the 1-$\sigma$ range found
from the fit to the 250-450~s curve.
The best-fit index $a$ ($\sim$0.75) is slightly larger than the value
$a = {2\over 3}$ expected for a spherical trapped fireball.
The larger fireball index could result
from a slight amount of neutrino cooling, or a negative temperature
gradient within the fireball.  By contrast, a deviation from
spherical geometry would tend to {\it decrease} $a$.
A large excess flux during the first $\sim 40$ s disappears just
as the large amplitude pulsations emerge.
Panel (b) shows on a linear scale how well the trapped fireball model
fits the final drop of the light curve, which cannot be described by
an exponential decay (see Fig.~\ref{3decays}).

\begin{figure}
\figurenum{2}
\plotone{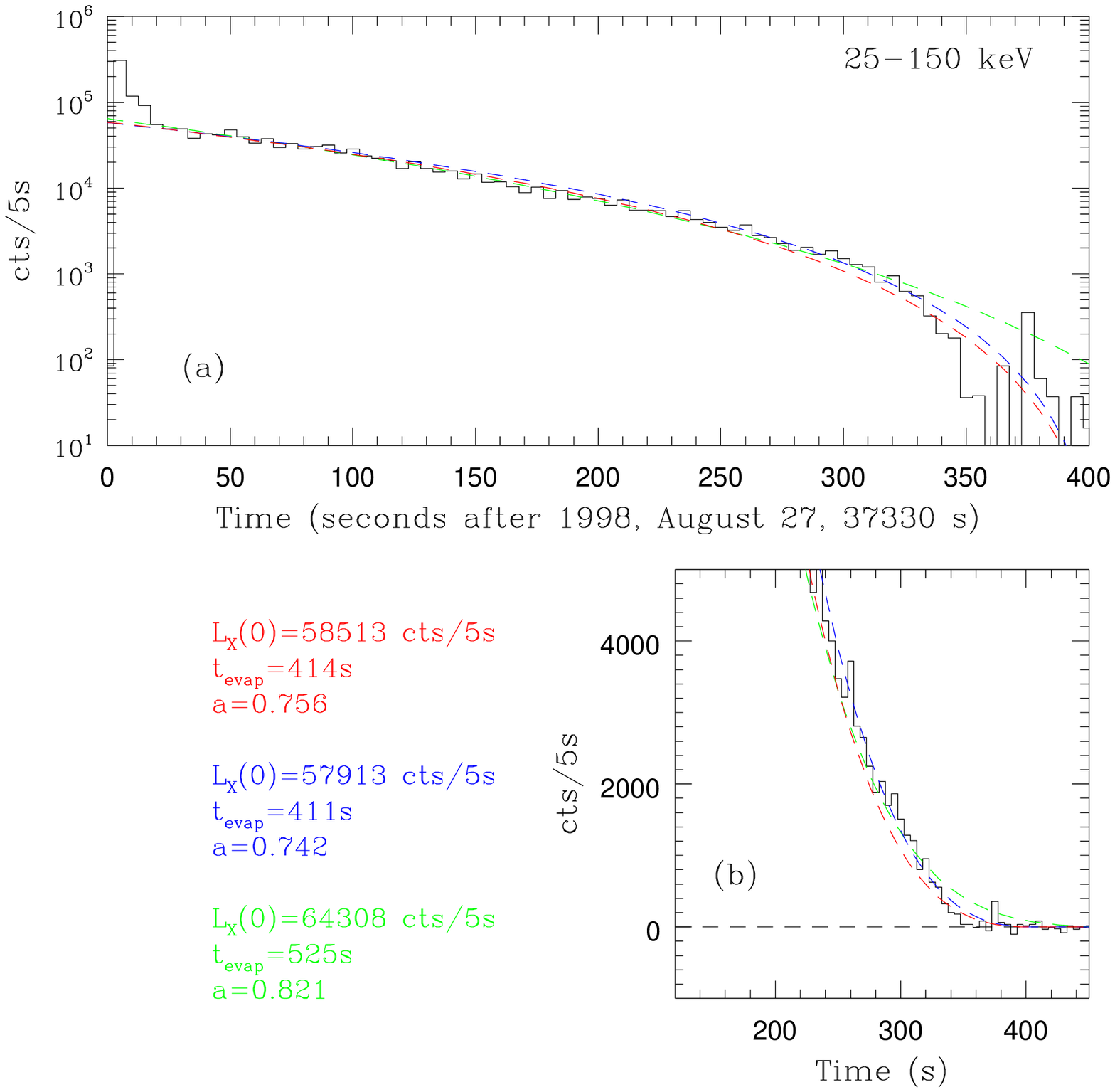}
\caption{
}
\label{fire_ul_reb}
\end{figure}

\begin{figure}
\figurenum{3}
\plotone{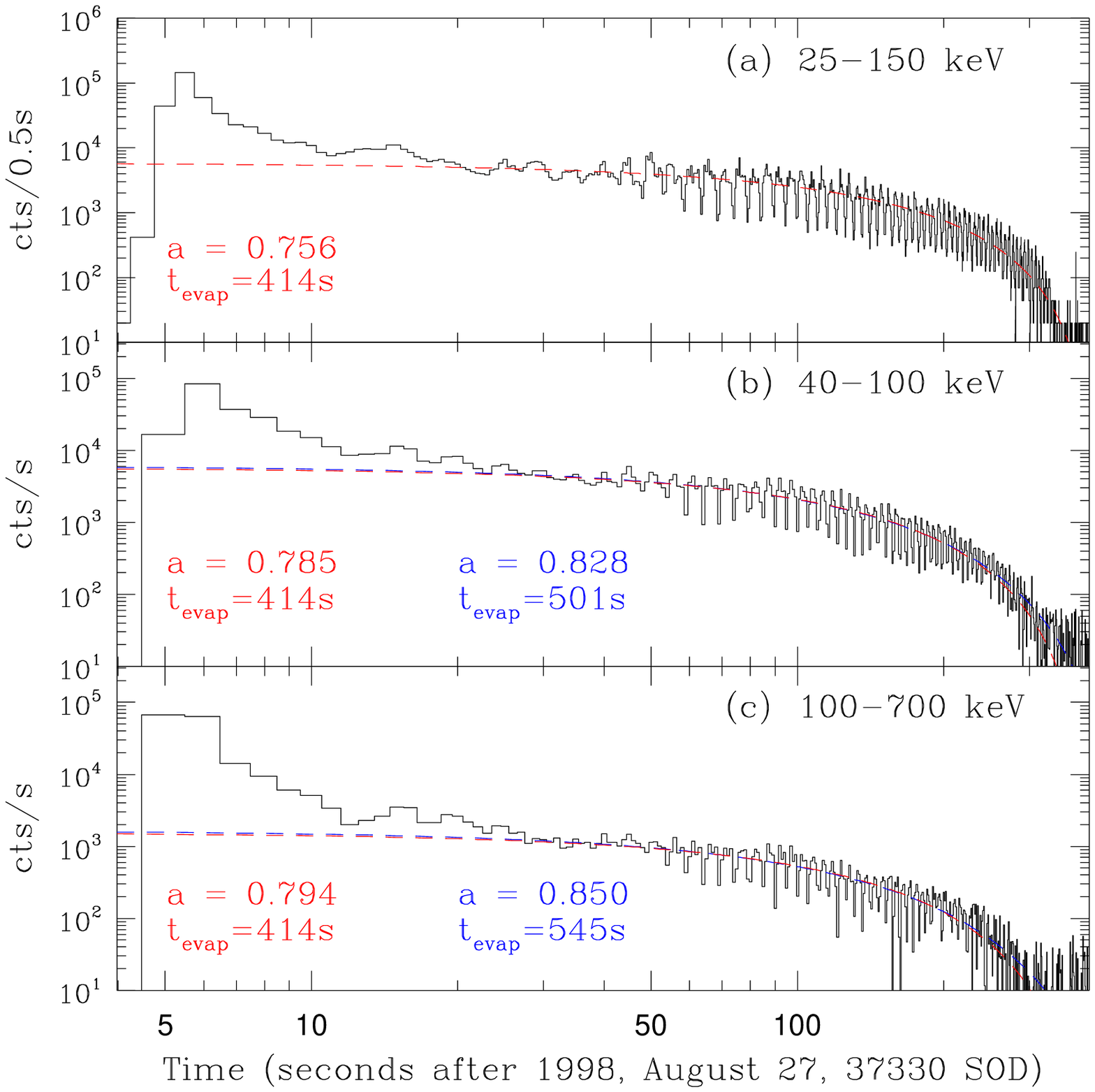}
\caption{
}
\label{fire_3decays}
\end{figure}

\begin{figure}
\epsscale{1}
\figurenum{4}
\plotone{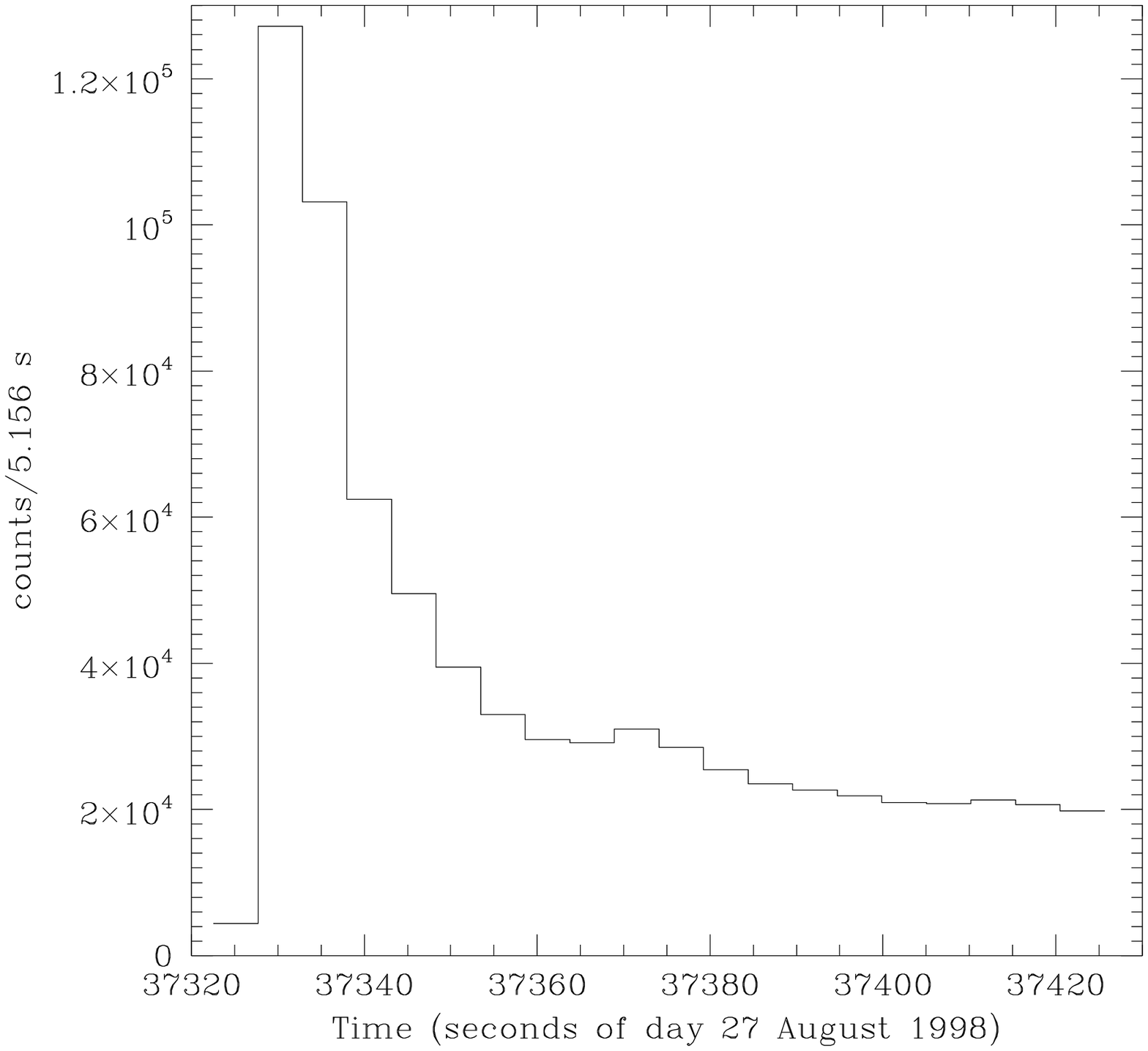}
\caption{
}
\label{smooth_decay}
\end{figure}

\begin{figure}
\figurenum{5}
\plotone{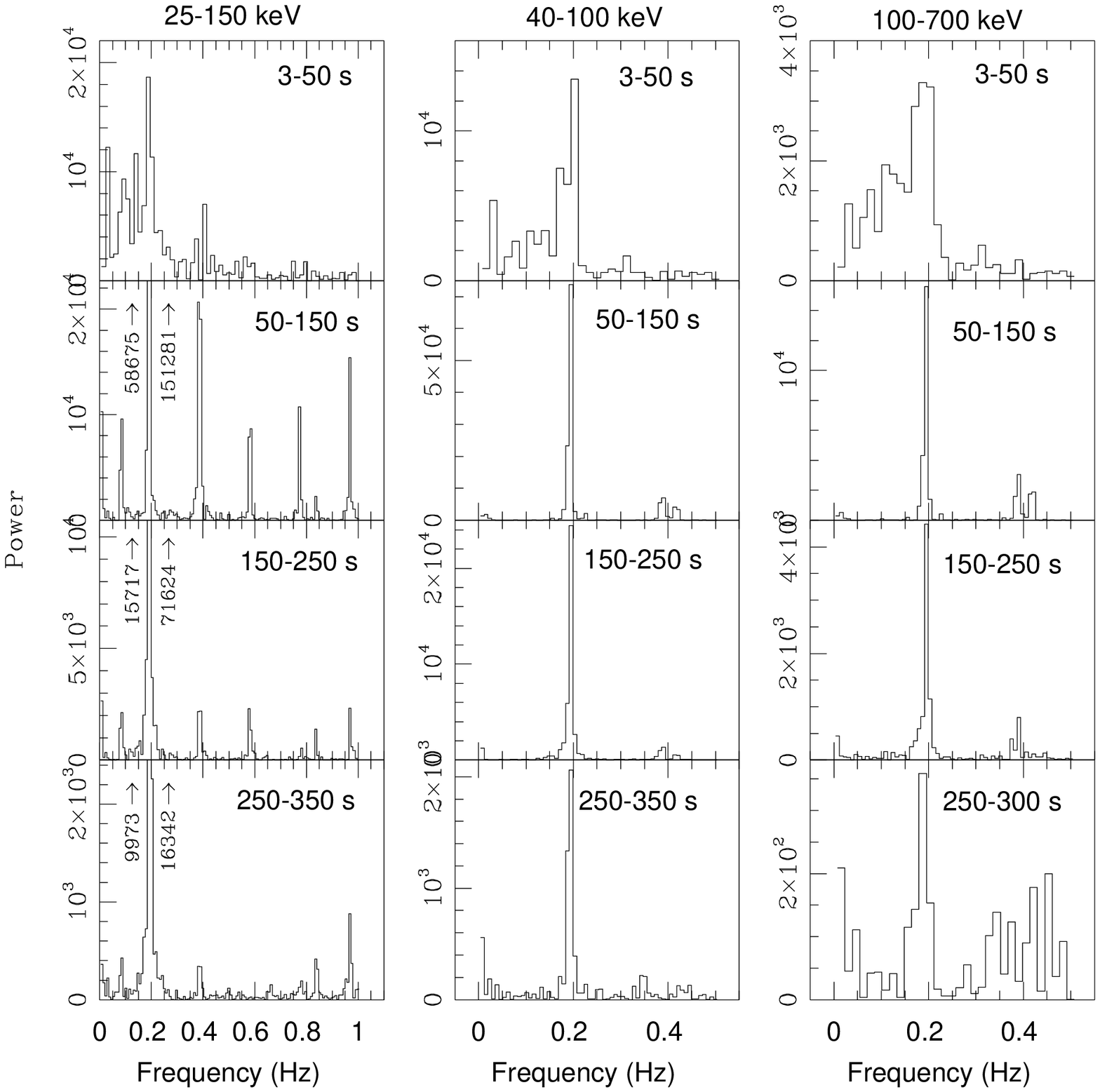}
\caption{
}
\label{fig_psd}
\end{figure}

\begin{figure}
\figurenum{6}
\plotone{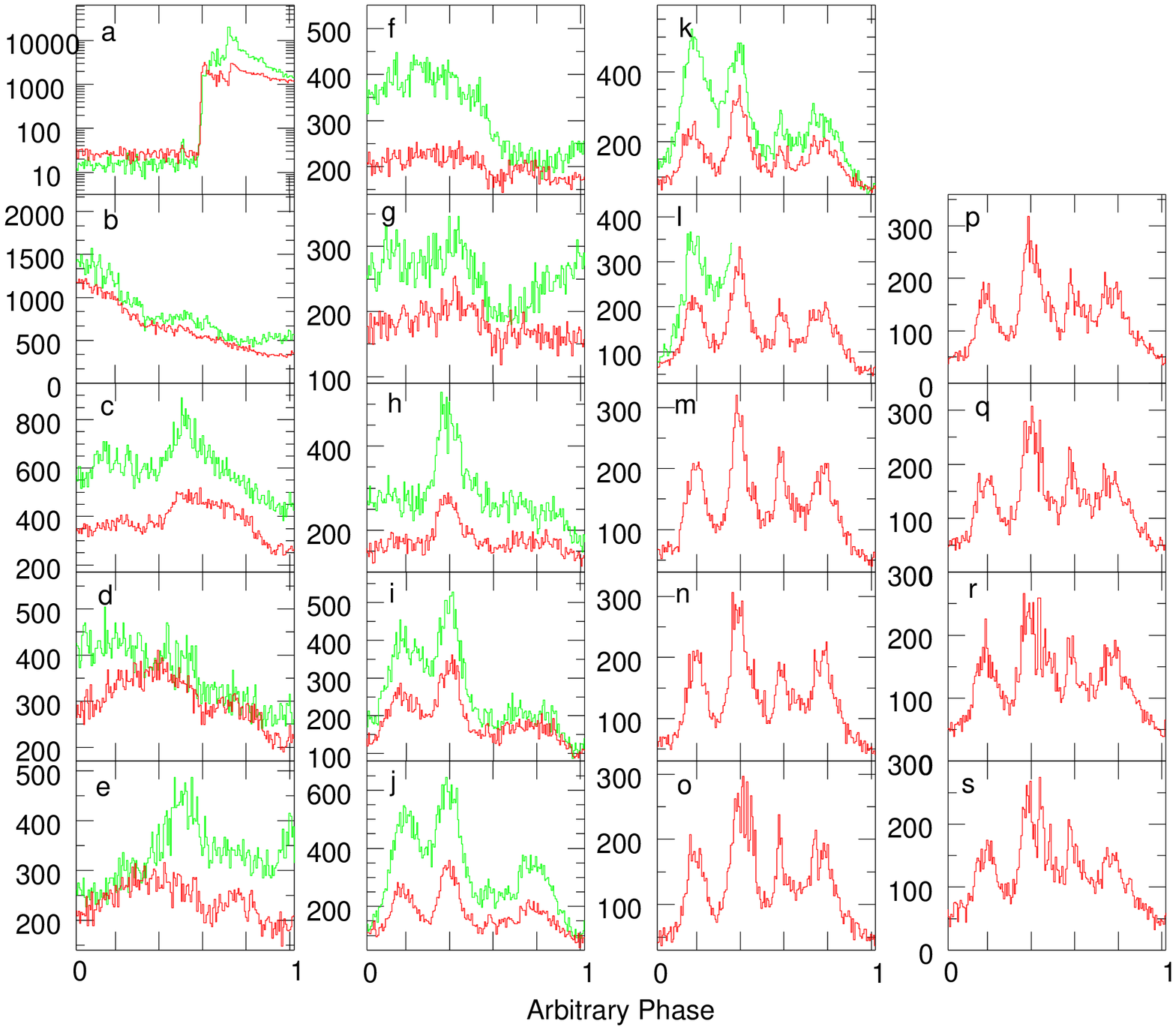}
\caption{
}
\label{grbm_panels}
\end{figure}

\begin{figure}
\epsscale{0.8}
\figurenum{7}
\plotone{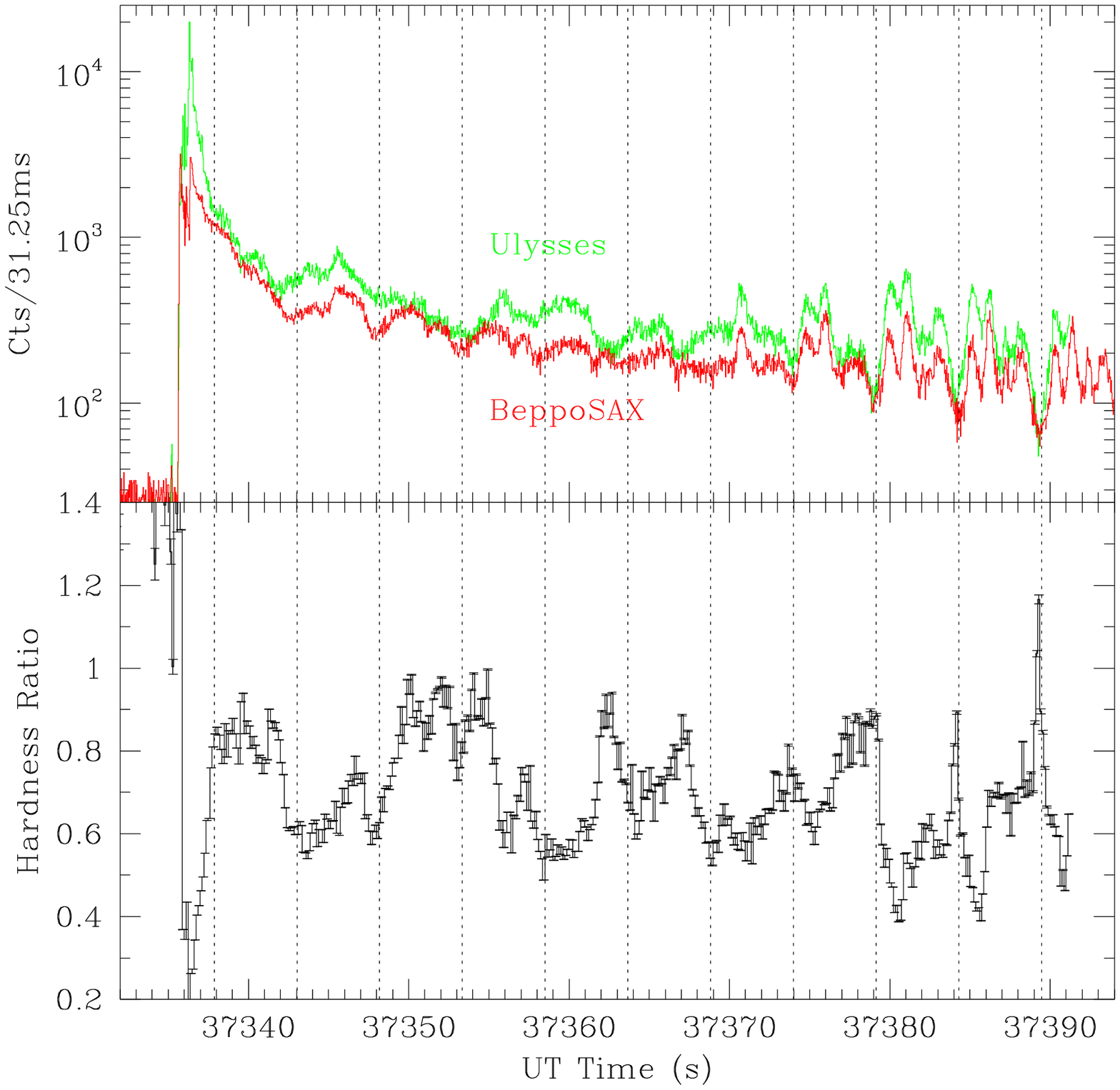}
\caption{
}
\label{Ulysses_grbm_diff}
\end{figure}

\begin{figure}
\figurenum{8}
\plotone{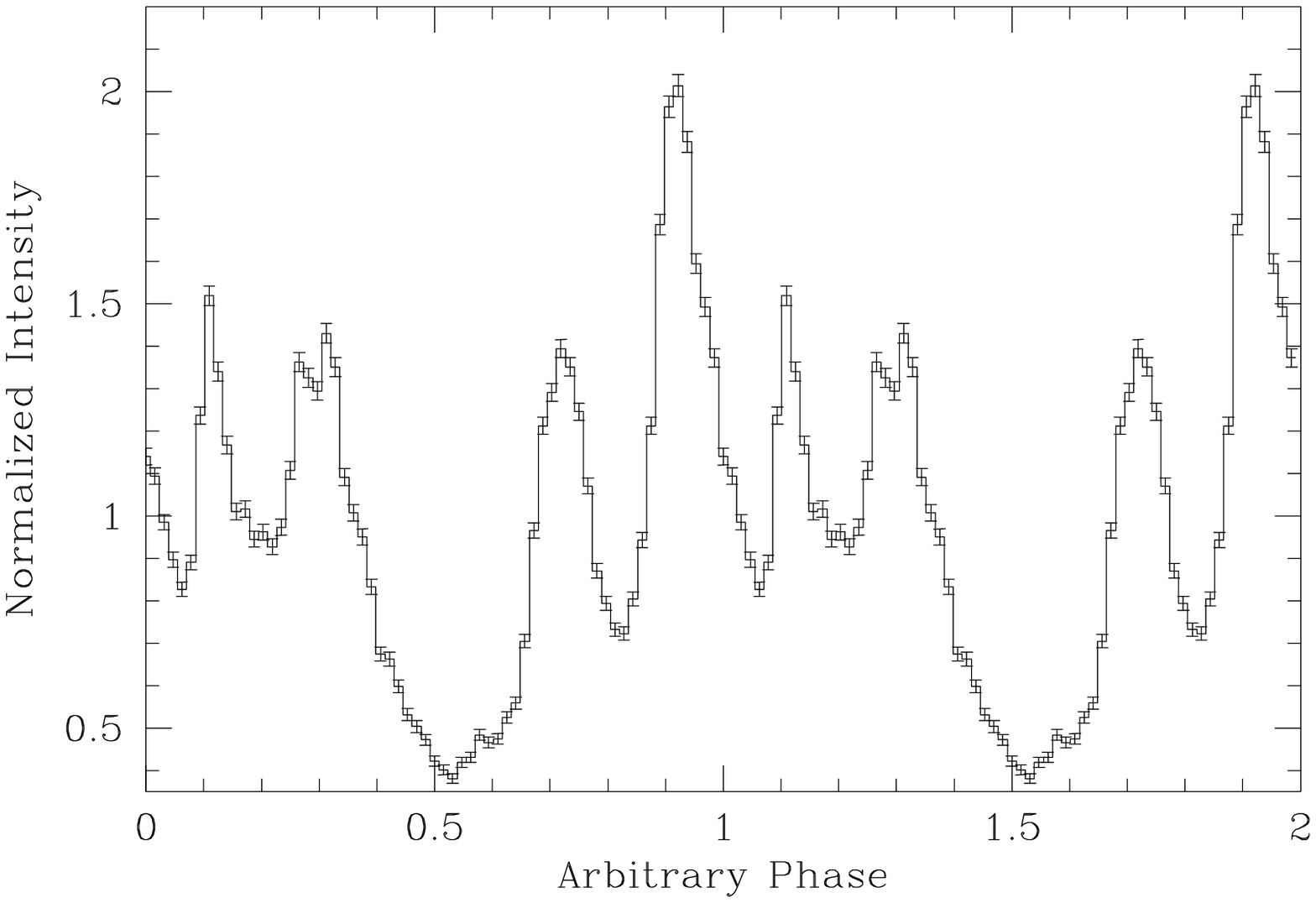}
\caption{
}
\label{folded_grbm}
\end{figure}

\begin{figure}
\figurenum{9}
\plotone{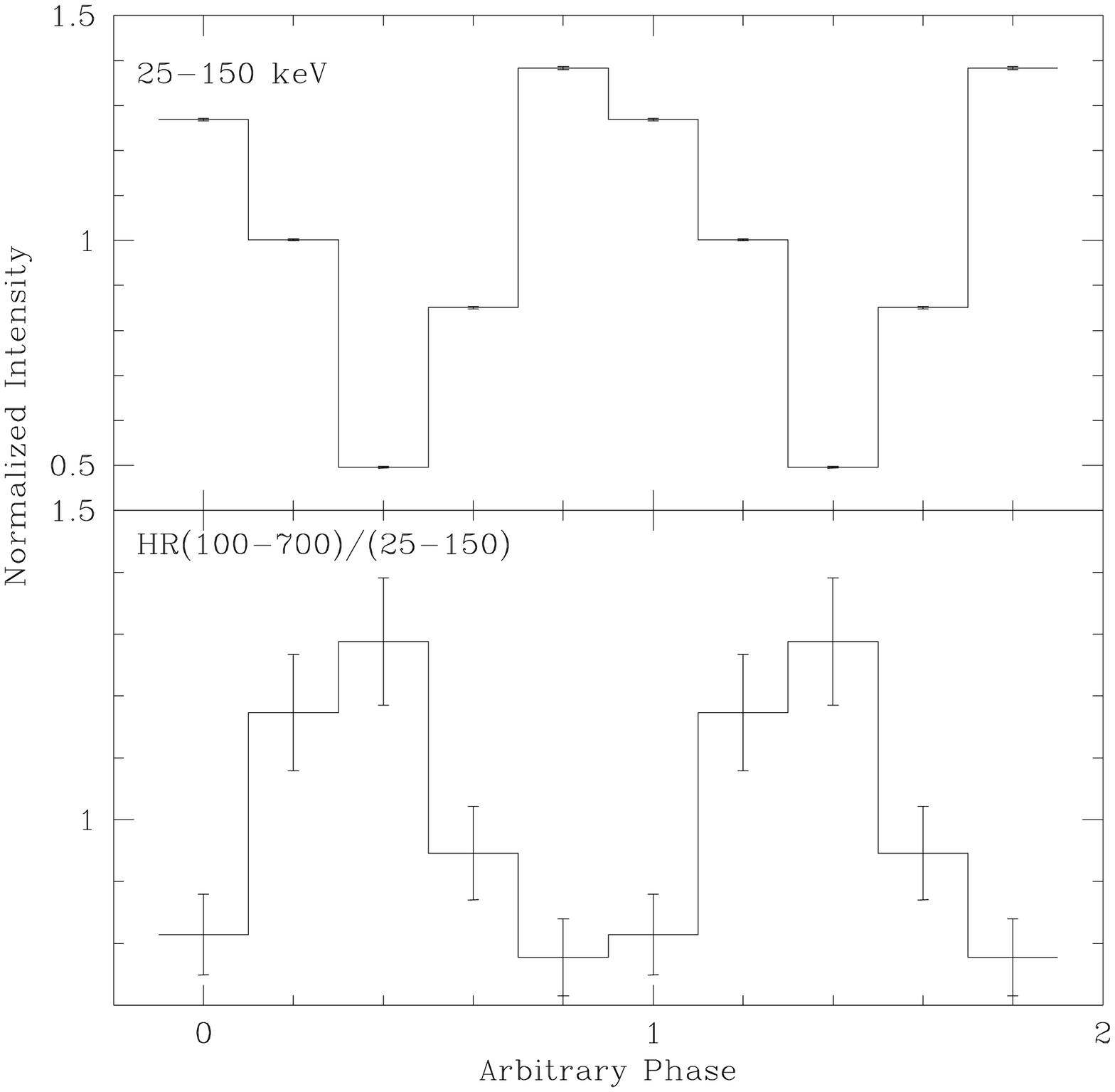}
\caption{
}
\label{folded_ul}
\end{figure}

\begin{figure}
\figurenum{10}
\plotone{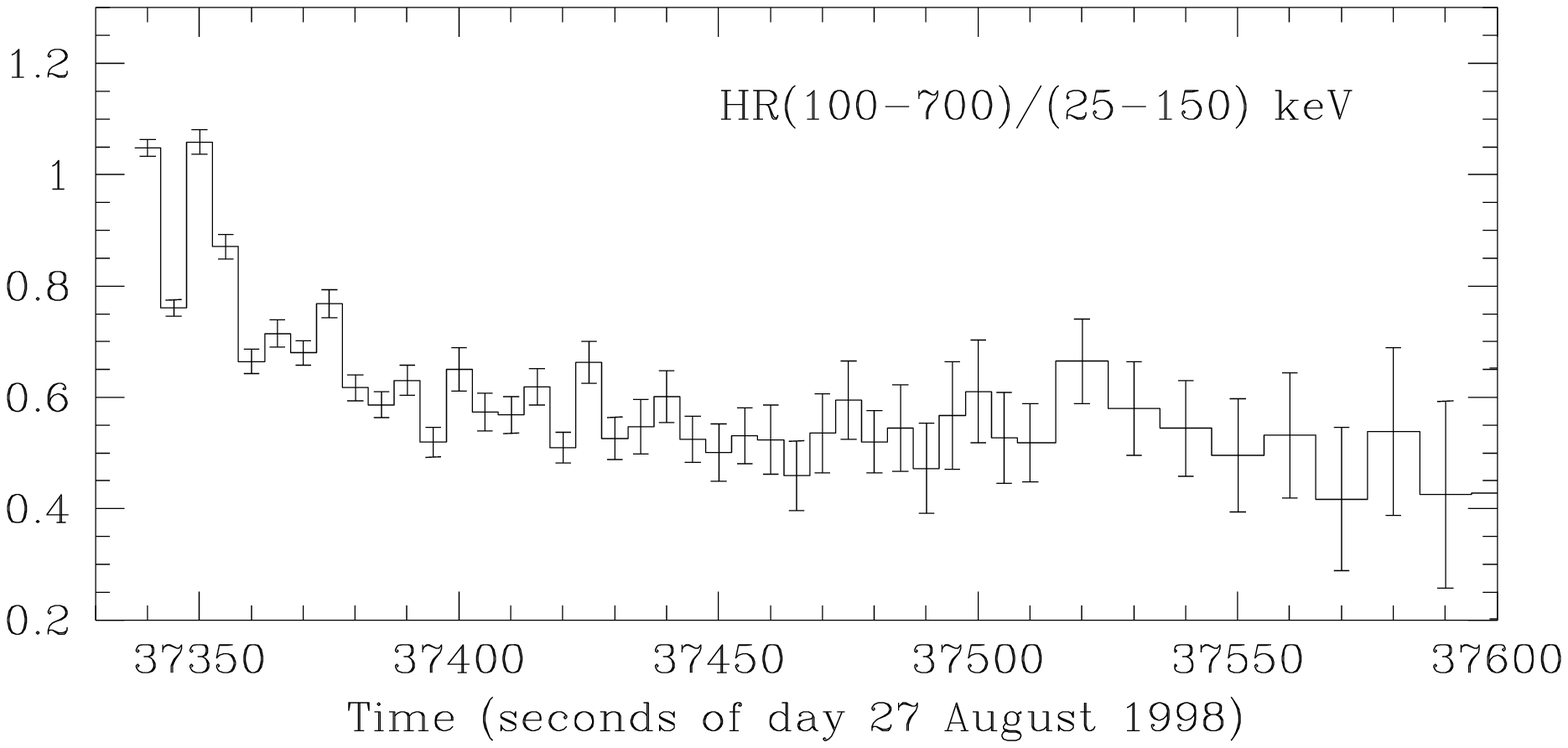}
\caption{
}
\label{hr}
\end{figure}

FIGURE 3:
\it Panel (a): \rm
The same analytical model as in Fig.~\ref{fire_ul_reb} (red curve) is
shown against the 0.5~s \it Ulysses \rm lightcurve.
\it Panel (b): \rm
The same trapped fireball model is fit to the 40-100~keV GRBM 1-s lightcurve. 
The red dashed line was obtained constraining the $t_{evap}$ parameter to
vary within the 1-$\sigma$ range found for 25-150~keV, allowing only the
$L_{X}(0)$ and $a$ to vary freely. The blue dashed curve is the best-fit,
without any constraint to the parameters.
\it Panel (c): \rm
Same as panel (b), but for the GRBM 100-700~keV 1-s lightcurve.
Note that here the initial excess appears significantly larger, relative
to the fireball model, than it does in the 40-100 keV light
curve.  This indicates a softening of the spectrum during the first 40 s.

FIGURE 4:
BeppoSAX GRBM 40--700 keV high time-resolution light curve, rebinned to
5.15625 s, which is our best approximation to the stellar rotation period.
This shows that the cycle-averaged emission decays nearly smoothly and
monotonically.

FIGURE 5:
Power Spectral Densities of the August 27 light curve
in four consecutive time intervals (four rows), using \it Ulysses \rm data
with 0.5~s time resolution (first column),  and BeppoSAX data with 1~s
time resolution, in two GRBM energy ranges (second and third columns).
The reference time is the onset of the event: 37335.168 SOD for
\it Ulysses \rm and 37335.0565 SOD for GRBM.
The time intervals, with respect to the reference times,
chosen to perform the FFTs are indicated in the panels.
In the high-energy band (100--700~keV) beyond 300 s the statistical quality
of the data did not allow to perform an FFT.
The data were detrended using the best-fit fireball models for each
energy range. For the first 50~s an additional second-order
polynomial detrending procedure was applied to the data to account
for the excess in the flux.
The bottom three panels for 25--150~keV have a vertical scale limited at
a 15\% of the maximum value (reached in two bins around 0.2~Hz and
indicated with an arrow).

FIGURE 6:
Pulse shape evolution from the BeppoSAX 40--700~keV (red) and \it
Ulysses \rm 25--150~keV (green) 32~ms resolution data.
Horizontal scale gives the phase of the stellar rotation cycle,
with zero phase taken at the deep minimum.
Each tickmark on the horizontal scale is 1/5 cycle, or 1.032~s.
Vertical scale gives counts per 32 ms.

FIGURE 7:
BeppoSAX 40--700~keV (red) and \it Ulysses \rm
25--150~keV (green) 31.25 ms count rates (top panel),
and their ratio (bottom panel).
Vertical dotted lines are spaced by one 5.16~s period.

FIGURE 8:
Folded light curve of the last eight 5.16 s pulses of the August 27th
event observed with the GRBM in the 40--700~keV range.

FIGURE 9:
\it Ulysses \rm (25-150 keV) light curve and
hardness ratio (100-700 keV)/(25-150 keV)
for the August 27th event from SGR 1900+14,
folded at an arbitrary epoch and with a period of 5.1589715~s.
This plot includes data beginning 50 s after the flare's onset (after
the four-peaked pattern was well-established) and extending to the
end of the event.

FIGURE 10:
Hardness ratio between GRBM (100--700~keV) and
\it Ulysses \rm (25-150 keV) light curves,
rebinned to 5~s in order to average (approximately) over rotation cycles
and show the general spectral trend. After t$\sim$37510~s the average is
performed over 10~s to compensate for the lower statistics.

FIGURE 11:
\it Top \rm:
Combined \it Ulysses \rm and GRBM energy spectra of the second time
interval (Interval B) and the best fitting spectral model.
The spectral model is composed by two blackbody models ($kT=9.3$ and
20.2~keV, respectively) and a power law (photon index 2.8).
\it Bottom \rm:
Combined \it Ulysses \rm and GRBM energy spectra of the third time
interval (Interval C)
and its best fit with an Optically Thin Thermal Bremmstrahlung spectral
model ($kT=$26.3 keV).

FIGURE 12:
March 5, 1979 light curve in two energy ranges, as measured during the
first 32 seconds of the event by the SIGNE experiment onboard the
Venera 12 spacecraft.

FIGURE 13:
March 5, 1979 light curve in the 130-353~keV nominal energy range
(Top Panel) and hardness ratio of counts in 203-353~keV to counts
in 130-203~keV (Bottom Panel), as measured during the
first 32 seconds of the event by the SIGNE experiment onboard the
Venera 11 and 12 spacecraft.  Data were folded on the 8-s rotation period
of the star. The characteristic pulse/interpulse pattern of the
March 5th light curve is apparent in the top panel.
Counts were corrected for systematics due to onboard compression.
Error bars were derived by Monte Carlo simulations.

FIGURE 14:
\it Panel (a) \rm:
The ISEE-3 data (above 50~keV, Cline et al. 1982)
of the March 5, 1979 event is compared with
exponential decay laws with time-constants $\tau = 60$ and 80 s;
and with the best-fit parameterization of a cooling, trapped
fireball (eq. [\ref{lxt}]). The event trigger is at $t$=5~s on the
horizontal scale.
\it Panel (b) \rm:
Comparison of the ISEE-3 data, rebinned to 8-s intervals, with the same
analytic models.
The rapid drop in flux after $\sim 60$ s appears to be better-fit
by models of emission from the contracting boundary of a trapped fireball
than by exponential decay models.

\newpage

\clearpage
\begin{deluxetable}{ccccc}
\tablecaption{\it BeppoSAX GRBM/\it Ulysses \rm  energy spectra
integration times.}
\tablehead{
 &  \it Ulysses  & \it Ulysses \rm & GRBM  & GRBM\\
\colhead{\it Interval} & \colhead{UT Start Times, s} & \colhead{UT End
Times, s}
 & \colhead{UT Start Times, s} & \colhead{UT End Times, s}
}
\startdata

A &   37335.268    & 37398.235  & 37335.7         &       37402.7 \\
B &   37398.235    & 37526.238  & 37402.7         &       37530.7 \\
C &   37526.238    & 37654.238  & 37530.7         &       37658.7 \\

\enddata
\end{deluxetable}

\begin{deluxetable}{cccccccc}
\tablecaption{\it BeppoSAX GRBM/\it Ulysses \rm  energy spectra fitting
parameters for interval B. The tested models are superpositions of
the following individual components: blackbody (BB), optically thin
thermal bremsstrahlung (OTTB), bremsstrahlung (BREMSS), power law
(PL) and the Band function (BAND, Band et al. 1993) }
\tablehead{
\colhead{\it Model} & \colhead{\it kT1} & \colhead{\it kT2} &
\colhead{$\alpha$}
& \colhead{$\beta$} & \colhead{$E_{0}$}  & \colhead{reduced $\chi2$}
& \colhead{GRBM/\it Ulysses \rm normalization }
}
\startdata

BB+BB+PL    &  9.3 keV  & 20.2 keV  & 2.8 & -   & -        & 1.62 (81
dof) & 0.43 \\
OTTB+PL     & 22.7 keV  & -         & 3.0 & -   & -        & 2.43 (82
dof) & 0.44 \\
BREMSS+PL   & 27.0 keV  & -         & 2.7 & -   & -        & 2.62 (82
dof) & 0.43 \\
BB+PL       & 10.8 keV  & -         & 3.4 & -   & -        & 3.40 (82
dof) & 0.45 \\
BAND        & -         & -         & 2.3 & 5.4 & 44.9 keV & 3.95 (82
dof) & 0.38 \\

\enddata
\end{deluxetable}

\begin{deluxetable}{cccccccc}
\tablecaption{\it BeppoSAX GRBM/\it Ulysses \rm  energy spectra fitting
parameters for interval C. Models are the same as Table 2, except for
the cut-off power law (CUTOFFPL).}
\tablehead{
\colhead{\it Model} & \colhead{\it kT1} & \colhead{\it kT2} &
\colhead{$\alpha$}
& \colhead{$\beta$} & \colhead{$E_{0}$}  & \colhead{reduced $\chi2$}
& \colhead{GRBM/\it Ulysses \rm normalization }
}
\startdata

CUTOFFPL    & 33.1 keV  & -         & 1.6 & -   & -      & 1.004 (64
dof) & 0.34 \\
BREMSS      & 30.7 keV  & -         & -   & -   & -      & 1.005 (65
dof) & 0.33 \\
BB+PL       & 12.2 keV  & -         & 3.6 & -   & -      & 1.051 (63
dof) & 0.33 \\
OTTB        & 26.3 keV  & -         & -   & -   & -      & 1.081 (65
dof) & 0.32 \\
BAND        & -         & -         & 2.3 & 8.9 & 50 keV & 1.119 (63
dof) & 0.35 \\
BB          & 14.0 keV  & -         & -   & -   & -      & 2.108 (68
dof) & 0.31 \\

\enddata
\end{deluxetable}

\newpage

\clearpage

\clearpage
\clearpage

\clearpage

\clearpage

\clearpage

\clearpage

\clearpage

\clearpage

\clearpage

\clearpage
\begin{figure}
\figurenum{11 Top}
\plotone{feroci11top.ps}
\caption{
}
\label{spectra_a}
\end{figure}

\clearpage
\begin{figure}
\figurenum{11 Bottom}
\plotone{feroci11bot.ps}
\caption{
}
\label{spectra_b}
\end{figure}

\clearpage
\begin{figure}
\figurenum{12}
\plotone{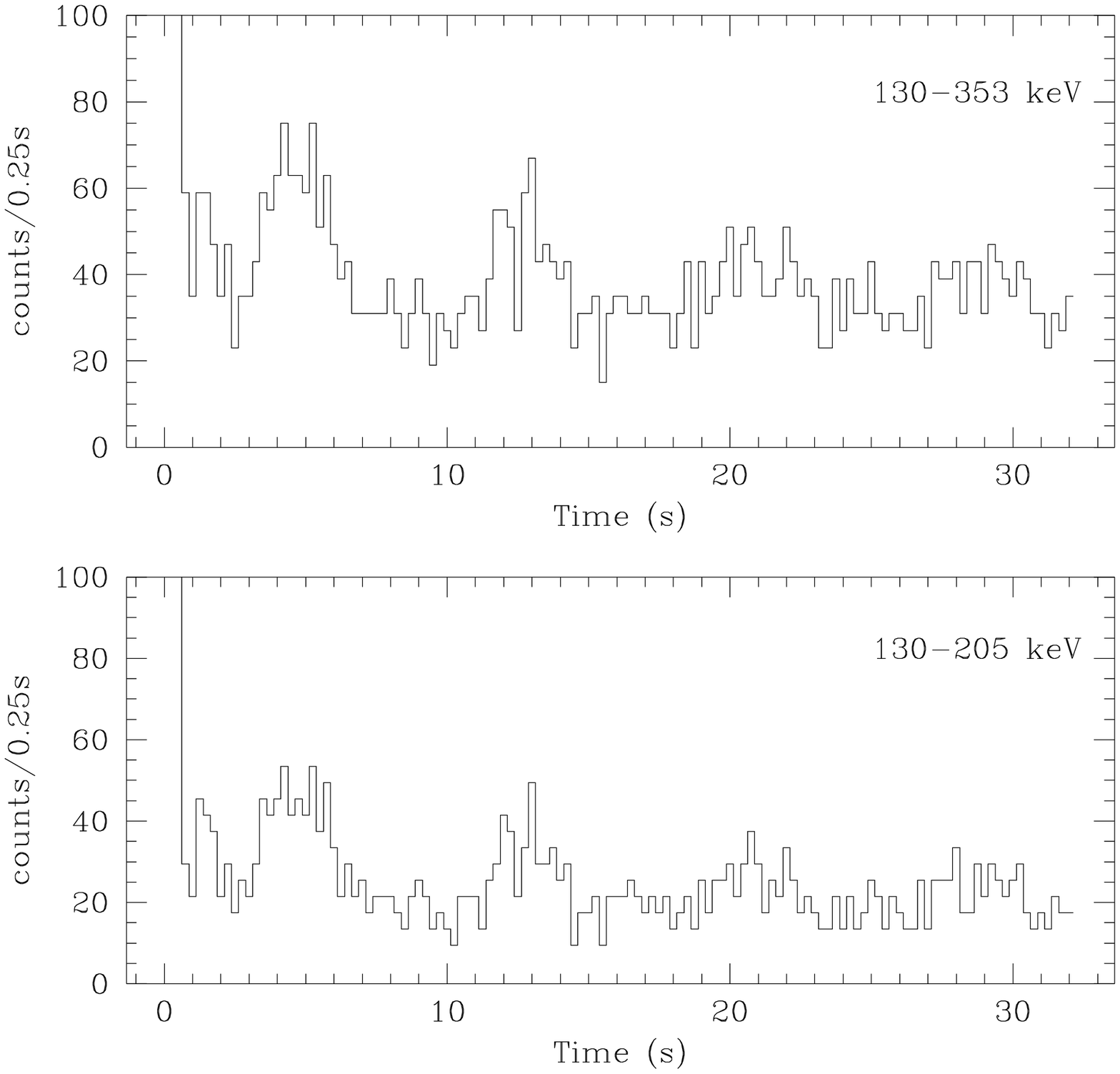}
\caption{
}
\label{signelc}
\end{figure}

\clearpage
\begin{figure}
\epsscale{1}
\figurenum{13}
\plotone{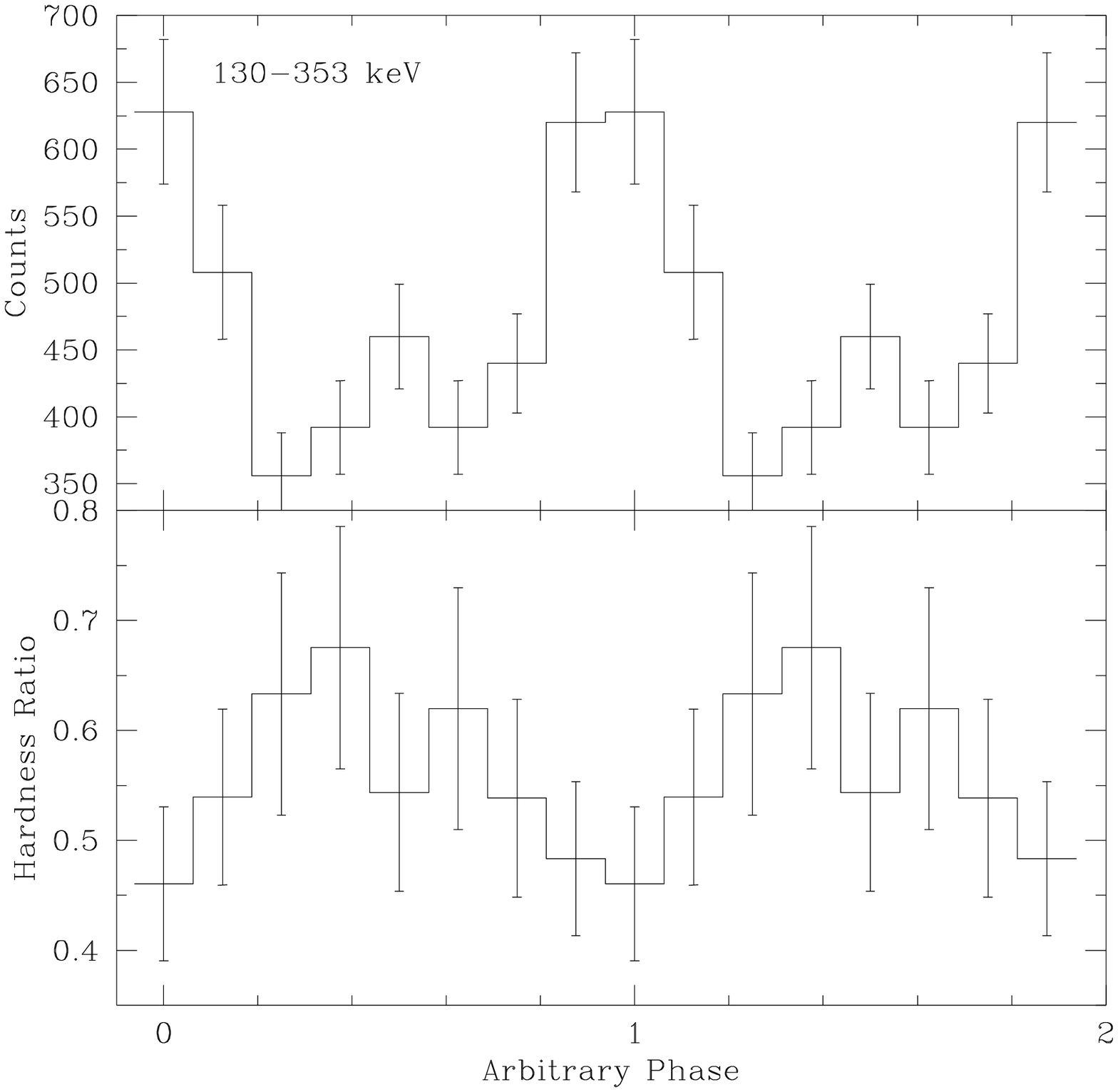}
\caption{
}
\label{venera_march5}
\end{figure}

\clearpage
\begin{figure}
\figurenum{14}
\plotone{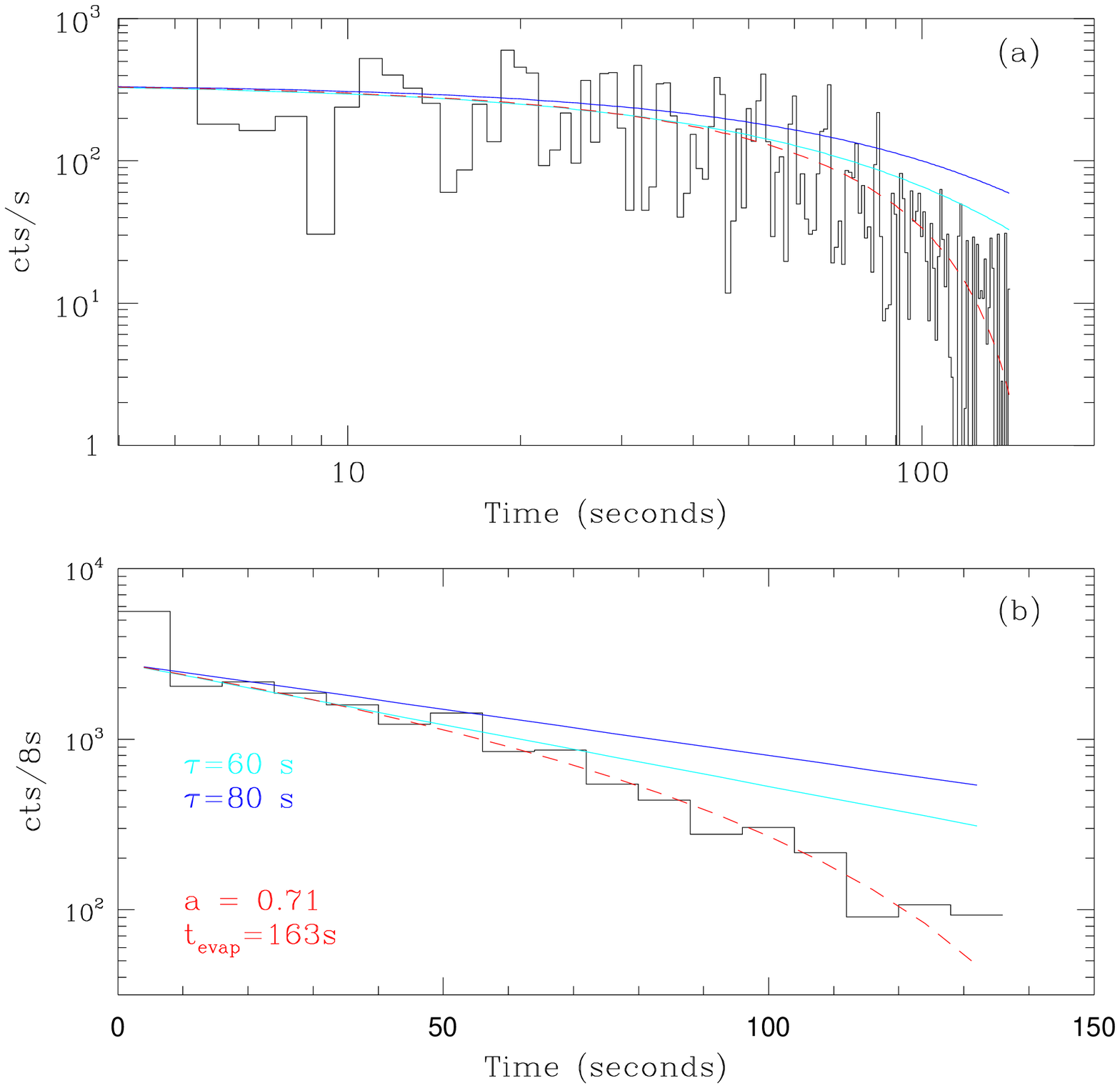}
\caption{
}
\label{fire_venera}
\end{figure}


\begin{thebibliography}{aa}

\bibitem[Alpar 2000]{alpar00}
Alpar, M.A. 2000, preprint, astro-ph/9912228

\bibitem[Amati et~al.\ 1997]{amati97}
Amati, L. et~al. 1997, SPIE Proceedings, 3114, 176

\bibitem[Amati\ 1999]{amati99}
Amati, L. 1999, PhD Thesis, University of Rome "La Sapienza"
(\it http://tonno.tesre.bo.cnr.it/$\sim$amati/tesi/ \rm).

\bibitem[Band et~al. 1993]{band93}
Band, D. et~al. 1993, ApJ, 413, 281

\bibitem[Barat et~al. 1979]{barat79}
Barat, C., Chambon, G., Hurley, K., Niel, M., Vedrenne, G.,
Estuline, I.V., Kurt, V.G., and Zenchenko, V.M. 1979, \aap, 79, L24

\bibitem[Barat et~al. 1981]{barat81}
Barat, C., Chambon, G., Hurley, K., Niel, M., Vedrenne, G., Estulin,
I.V., Kuznetsov, A.V., and Zenchenko, V.M. 1981, Space Sci. Inst., 5,
229

\bibitem[Barat et~al. 1983]{barat83}
Barat, C., Hayles, R.I., Kurley, K., Niel, M., Vedrenne, G., Desai, U.,
Estulin, I.V., Kurt, V.G., \& Zenchenko, V.M. 1983, \aap, 126, 400

\bibitem[Blaes et al. 1989]{blaes89}
Blaes, O., Blandford, R.D., Goldreich, P., \& Madau P. 1989, \apj,
343, 829

\bibitem[Chatterjee, Hernquist, \& Narayan 2000]{chatterjee00}
Chatterjee, P., Hernquist, L. \& Narayan, R. 2000, ApJ, 534, 373 

\bibitem[Cheng et al.~1996]{cheng96}
Cheng, B., Epstein, R.I., Guyer, R.A. \& Young, C. 1996, Nature, 382,
518

\bibitem[Cline et~al. 1998]{cline98}
Cline, T., Mazets, E. and Golenetskii, S.V., 1998, IAU Circular no. 7002 

\bibitem[Cline et~al. 1982]{cline82}
Cline, T. et al. 1982, \apj \, 255, L45

\bibitem[Duncan \& Thompson 1992]{duncan92}
Duncan, R.C. \& Thompson, C., 1992, ApJ, 392, L9 (DT92)

\bibitem[Duncan 1998]{duncan98}
Duncan, R.C. 1998, ApJ, 498, L45

\bibitem[Feroci et~al.\ 1997]{feroci97}
Feroci, M. et~al. 1997, SPIE Proceedings, 3114, 186

\bibitem[Feroci et~al.\ 1998]{feroci98}
Feroci, M. et~al. 1998, IAU Circular no. 6945

\bibitem[Feroci et~al.\ 1999]{feroci99}
Feroci, M. et~al. 1999, ApJ 515, L9

\bibitem[Frail, Kulkarni and Bloom\ 1999]{frail99}
Frail, D., Kulkarni, S., and Bloom, J., 1999, Nature 398, 127

\bibitem[Frontera et~al.\ 1997]{frontera97}
Frontera, F., Costa, E., Dal~Fiume, D., Feroci, M., Nicastro, L.,
Orlandini,
M., Palazzi, E., and Zavattini, G. 1997, \AAS, 122, 357

\bibitem[Golenetskii et~al.\ 1987]{golenetskii87}
Golenetskii,S.V. et~al. 1987, Sov. Astron. Lett. 13(3), 166

\bibitem[Goldreich \& Reisenegger 1992]{goldreich92}
Goldreich, P. \& Reisenegger A. 1992, \apj, 395, 250

\bibitem[Gogus et al.~1999]{gogus99}
Gogus, E. et al.~1999, ApJ, 526, L93 

\bibitem[Gogus et al.~2000]{gogus00}
Gogus, E. et al.~2000, ApJ, 532, L121 

\bibitem[Gotthelf, Vasisht \& Dotani 1999]{gotthelf99}
Gotthelf, E.V., Vasisht, G. \& Dotani, T. 1999, ApJ, 522, L33

\bibitem[Harding, Contopoulous \& Kazanas 1999]{harding99}
Harding, A.K., Contopoulous, I. \& Kazanas, D. 1999, ApJ, 525, L125

\bibitem[Heyl and Hernquist 1997]{heyl97}
Heyl, J.S. \& Hernquist, L. 1997, ApJ, 489, L67

\bibitem[Heyl and Kulkarni 1998]{heyl98}
Heyl, J.S. \& Kulkarni, S.R. 1998, ApJ, 506, L61

\bibitem[Herold 1979]{herold79}
Herold, H. 1979, Phys.~Rev.~D, 19, 2668

\bibitem[Hurley et~al.\ 1992]{hurley92}
Hurley, K., et al. 1992, Astron. Astrophys. Suppl. Ser., 92(2), 401

\bibitem[Hurley et~al.\ 1999a]{hurley99b}
Hurley, K. et~al. 1999a, ApJ, 510, L107

\bibitem[Hurley et~al.\ 1999b]{hurley99a}
Hurley, K. et~al. 1999b, Nature 397, 41

\bibitem[Hurley et~al.\ 1999c]{hurley99c}
Hurley, K. et~al. 1999c, ApJ, 510, L111

\bibitem[Hurley et~al.\ 1999d]{hurley99d}
Hurley, K. et~al. 1999d, \apj \, 523, L37

\bibitem[Hurley et~al.\ 1999e]{hurley99e}
Hurley, K., Kouveliotou, C., Woods, P., Mazets, E., Golenetskii, S.,
Frederiks, D., Cline, T., and van Paradijs, J. 1999e, \apj \, 519, L143

\bibitem[Hurley et~al.\ 2000]{hurley0}
Hurley, K. et~al. 2000, \apj \, 528, L21

\bibitem[Inan et~al.\ 1999]{inan99}
Inan, U., Lehtinen, N., Lev-Tov, S., Johnson, M., Bell, T., and Hurley, K.
~1999, Geophys. Res. Lett., 26(22), 3357

\bibitem[Kouveliotou et~al.\ 1993]{kouveliotou93}
Kouveliotou, C., et~al. 1993, Nature, 362, 728

\bibitem[Kouveliotou et~al.\ 1998a]{kouveliotou98a}
Kouveliotou, C., et al., 1998a, \nat \, 393, 235

\bibitem[Kouveliotou et~al.\ 1998b]{kouveliotou98b}
Kouveliotou, C. et~al. 1998b, IAU Circular no. 6929, June 3, 1998

\bibitem[Kouveliotou et~al.\ 1998c]{kouveliotou98c}
Kouveliotou, C. et~al. 1998c, IAU Circular no. 6944 

\bibitem[Kouveliotou et~al.\ 1999]{kouveliotou99}
Kouveliotou et~al. 1999, ApJ, 510, L115

\bibitem[Li 1999]{li99}
Li, X.D. 1999, ApJ, 520, 271

\bibitem[Kulkarni \& Frail 1993]{kulkarni93}
Kulkarni, S. \& Frail, D., 1993, Nature, 365, 33

\bibitem[Marsden, Rothschild \& Lingenfelter 1999]{marsden99}
Marsden, D., Rothschild, R.E, \& Lingenfelter, R.E. 1999, ApJ, 520, L107

\bibitem[Mazets et~al.\ 1979a]{mazets79a}
Mazets, E.P., Golenetskii, S.V., and Gur'yan, Yu.A. 1979a,
Soviet Astron. Lett., 5(No.6),343

\bibitem[Mazets et~al.\ 1979b]{mazets79b}
Mazets, E.P., et al., 1979b, Nature, 282, 587

\bibitem[Mazets et~al.\ 1982]{mazets82}
Mazets, E.P., Golenetskii, S.V., Gur'yan, Yu.A., Ilyinskii, V., 1982,
Astrophys. Space Sci. 84, 173

\bibitem[Mazets et~al.\ 1999]{mazets99}
Mazets, E.P., et al., 1999, Astron. Lett., 25(10), 635 

\bibitem[McDermott, Van Horn \& Hansen 1988]{mcdermott88}
McDermott, P.N., Van Horn, H.M. \& Hansen, C.J.  1988, ApJ, 325, 725

\bibitem[Miller 1995]{miller95}
Miller, C.M. 1995, ApJ, 448, L29

\bibitem[Murakami et~al.\ 1994]{murakami94}
Murakami, T.. et al., 1994, Nature, 36, 129

\bibitem[Murakami et~al.\ 1999]{murakami99}
Murakami, T. et~al. 1999, ApJ, 510, L122

\bibitem[Paczy\`nski 1992]{paczynski92}
Paczy\`nski, B., 1992, Acta Astron., 42, 145

\bibitem[Palmer ~1999]{palmer99}
Palmer, D., ~1999, ApJ, 512, L113 

\bibitem[Reisenegger \& Goldreich 1992]{reisenegger92}
Reisenegger A. \& Goldreich P. 1992, ApJ, 395, 240

\bibitem[Sholz 1990]{sholz90}
Sholz, C.H. 1990, The Mechanics of Earthquakes and Faulting (Cambridge:
Cambridge Univ.~Press)

\bibitem[Smith, Bradt \& Levine 1999]{smith99}
Smith, D.A., Bradt, H.V. and Levine, A.M., 1999, ApJ, 519, L147

\bibitem[Thompson \& Blaes 1998]{thompson98}
Thompson, C. \& Blaes, O. 1998, \prd, 57, 3219

\bibitem[Thompson \& Duncan 1993]{thompson93}
Thompson, C. \& Duncan, R.C. 1993, ApJ, 408, 194 (TD93)

\bibitem[Thompson \& Duncan 1995]{thompson95}
Thompson, C., \& Duncan, R.C., 1995, \MN, 275, 255 (TD95)

\bibitem[Thompson \& Duncan 1996]{thompson96}
Thompson, C. \& Duncan, R.C. 1996, ApJ, 473, 322 (TD96)

\bibitem[Thompson et~al.\ 1999]{thompson99}
Thompson, C., Duncan, R.C., Woods, P.M., Kouveliotou, C., Finger, M.H.,
\& van Paradijs, J. 1999, ApJ, in press.  \ {\rm astro-ph/9908096}

\bibitem[Thompson et~al.\ 1999]{thompson00}
Thompson, C., et al. 2000, \it submitted to ApJ \rm (Paper II)

\bibitem[van Paradijs, Taam \& van den Heuvel 1995]{vanparadijs95}
van Paradijs, J., Taam, R.E \& van den Heuvel E.P.J. 1995, A\& A, 299, L41 

\bibitem[Vasisht et~al.\ 1994]{vasisht94}
Vasisht, G., et al., 1994, ApJ, 431, L35

\bibitem[Vasisht \& Gotthelf 1997]{vasisht99}
Vasisht, G. \& Gotthelf, E.V. 1997, ApJ, 486, L129

\bibitem[Woods et~al.\ 1999a]{woods99a}
Woods, P., Kouveliotou, C., van Paradijs, J., Hurley, K., Kippen, R.,
Finger, M.H., Briggs, M., Dieters, S., and Fishman, G.
  1999a, ApJ, 519, L139

\bibitem[Woods et~al.\ 1999b]{woods99b}
Woods, P., Kouveliotou, C., van Paradijs, J., Finger, M.H. \& Thompson, C.
  1999b, ApJ, 518, L103

\bibitem[Woods et~al.\ 2000]{woods00}
Woods, P., Kouveliotou, C., Finger, M.H., Gogus, E., Scott, D.M.,
Dieters, S., Thompson, C., Duncan, R.C., Hurley, K., Strohmayer, T.,
Swank, J. \& Murakahmi, T. 2000, ApJ, 535, L55 

\end{thebibliography}
\end{document}